\documentclass{aa}
\usepackage{graphicx}
\usepackage{txfonts}
\usepackage{hyperref}
\usepackage{xcolor}
\definecolor{airforceblue}{rgb}{0.36, 0.54, 0.66}
\definecolor{mod}{rgb}{0.56, 0.54, 0.16}
\definecolor{modtbd}{rgb}{0.56, 0.24, 0.16}
\definecolor{cornellred}{rgb}{0.7, 0.11, 0.11}
\definecolor{cadmiumgreen}{rgb}{0.0, 0.42, 0.24}
\hypersetup{
     colorlinks   = true,
     citecolor    = airforceblue,
     urlcolor     = cadmiumgreen,
     linkcolor    = cornellred
}

\newcommand{\mods}[1]{{#1}}
\newcommand{\modsTwo}[1]{{#1}}
\newcommand{\modsThree}[1]{{#1}}
\newcommand{\modsL}[1]{{#1}}

\begin{document}

   \title{Signatures of quenching in dwarf galaxies in local galaxy clusters}

   \subtitle{A comparison of the galaxy populations in the Virgo and Fornax clusters}
\authorrunning{J.~Janz et al.}
\titlerunning{Dwarfs in Virgo and Fornax}

   \author{Joachim Janz
          \inst{1,2},
          Heikki Salo\inst{2},
          Alan H. Su\inst{2}, \and Aku Venhola\inst{2}
          }

   \institute{$^1$Finnish Centre of Astronomy with ESO (FINCA), Vesilinnantie 5, FI-20014 University of Turku, Finland;\\
         %\and
             $^2$Space Physics and Astronomy Research Unit, University of Oulu, P.O. Box 3000, FI-90014 University of Oulu, Finland\\ 
             \email{\href{mailto:joachim.janz@oulu.fi}{joachim.janz@oulu.fi}}\\
             }

   \date{Received -- --, --; accepted -- --, --}
%September 15, 1996; accepted March 16, 1997
% \abstract{}{}{}{}{}
% 5 {} token are mandatory

  \abstract
   {The transformation of late-type galaxies has been suggested as the origin of early-type dwarf galaxies (typically $M_\star \le 10^9\, {\rm M_\odot}$) in galaxy clusters.    Based on deep images,  \mods{Venhola \modsL{and colleagues} analysed correlations between colour and surface brightness for galaxies in the Fornax cluster binned by luminosity or stellar mass. In the bins with $M_\star<10^8 {\rm M}_\odot$, the authors identified a correlation of redness with fainter surface brightness} and interpreted it as a consequence of the quenching of star formation by ram pressure stripping \mods{ in the dwarf galaxies}.
   }
   {  This study carries out a similar analysis  for the Virgo cluster. The analysis for both clusters 
    is then used to compare  \mods{the Virgo and Fornax clusters}, for which the ram pressure is expected to have different strengths.\mods{ The purpose of this is to scrutinise the ram pressure interpretation from the other study and search for differences between the clusters that reflect the different ram pressure efficiencies,
    which would either support or weaken this interpretation. Ultimately, this could help weigh the importance of ram pressure stripping 
    relative to other transformative processes in the shaping of the dominant early-type dwarf galaxy population.   }
   }
   {We extend the analysis of  colour versus surface brightness binned by stellar mass
   to higher masses and a wider range of optical colours. The results, in particular at low stellar mass,
   are compared to predictions of stellar evolution models. Benefitting from larger sample sizes, 
   we also analyse late- and early-type galaxies separately. This analysis is carried out for the Virgo
   and Fornax clusters, and the colour versus surface brightness relation, as well as other properties of the two clusters' galaxy populations, 
   are compared.}
   {While the colour--surface brightness diagrams are remarkably similar for the two clusters,
  only the low-mass late-type galaxies are found to have slopes consistent with
  a fading and reddening following the quenching of star formation. \modsTwo{For the early-type galaxies, there are no (or
  only weak) correlations between colour and surface 
  brightness in all mass bins.}
  Early- and late-type galaxies in both clusters have comparable sizes
  below a stellar mass of  $M_\star \lesssim 10^8\, {\rm M_\odot}$.
  The colour and size scaling relations are very similar
  for the Virgo and Fornax clusters. However, Virgo features a lower fraction of early-type
  or red galaxies despite its higher mass.}
   {The similarity of early-type dwarfs and low-mass late types \mods{in size at the masses
   $M_\star \lesssim 10^8\, {\rm M_\odot}$ as well as the overall consistency of
   the colour--surface brightness correlation with fading stellar populations}  support 
    a scenario of transformation \mods{via the quenching of star formation, \modsTwo{for example} by gas removal.}
   However, the lack of this imprint of an ageing stellar population on the early-type dwarfs 
   themselves calls for some additional explanation.
   Finally, the Virgo cluster is an atypical cluster with a comparably low fraction of quiescent
   early-type galaxies at all galaxy masses despite its large cluster mass. }

   \keywords{Galaxies: dwarf -- Galaxies: evolution -- Galaxies: stellar content -- Galaxies: clusters: general -- 
              Galaxies: clusters: individual: Virgo -- Galaxies: clusters: individual: Fornax }
   \maketitle
%
%-------------------------------------------------------------------
\section{Introduction}

\mods{Galaxy populations vary  from
low-density environments to massive galaxy clusters.
The morphology--density relation (\citealt{1980ApJ...236..351D}, see also e.g.~\citealt{2015MNRAS.451.3427H})
extends to lower-density environments, such as galaxy groups \citep[e.g.][]{1984ApJ...281...95P}, and 
is also observed at higher redshifts \citep[e.g.][]{2005ApJ...623..721P,2007ApJ...670..206V,2007ApJ...670..190H,2012ApJ...754..141M, 2013ApJ...770...58B}.
In the regime of dwarf galaxies (commonly defined as galaxies with $M_\star \le 10^9\, {\rm M_\odot}$; see, however,
\citealt{1987AJ.....94..251B}, whose classifications of Virgo galaxies into dwarf and normal galaxies follow a limit in surface brightness instead of luminosity or mass, as also reviewed in \citealt{2009ARA&A..47..371T}), it manifests itself with  galaxies with early-type morphology dominating 
the high-density cluster environments of massive galaxy clusters, 
while the low-density environments are  typically populated  
by more irregular late-type galaxies (ETGs; \citealt{1987AJ.....94..251B}, see also, e.g.~\citealt{2009AJ....138.1037C} for a lower-density group environment). While quiescent dwarf galaxies with $M_\star \le 10^9\, {\rm M_\odot}$ are common in galaxy clusters, they are nearly absent in the field \citep{2012ApJ...757...85G}.}

\begin{figure*}
  \centering
\includegraphics[width=\textwidth]{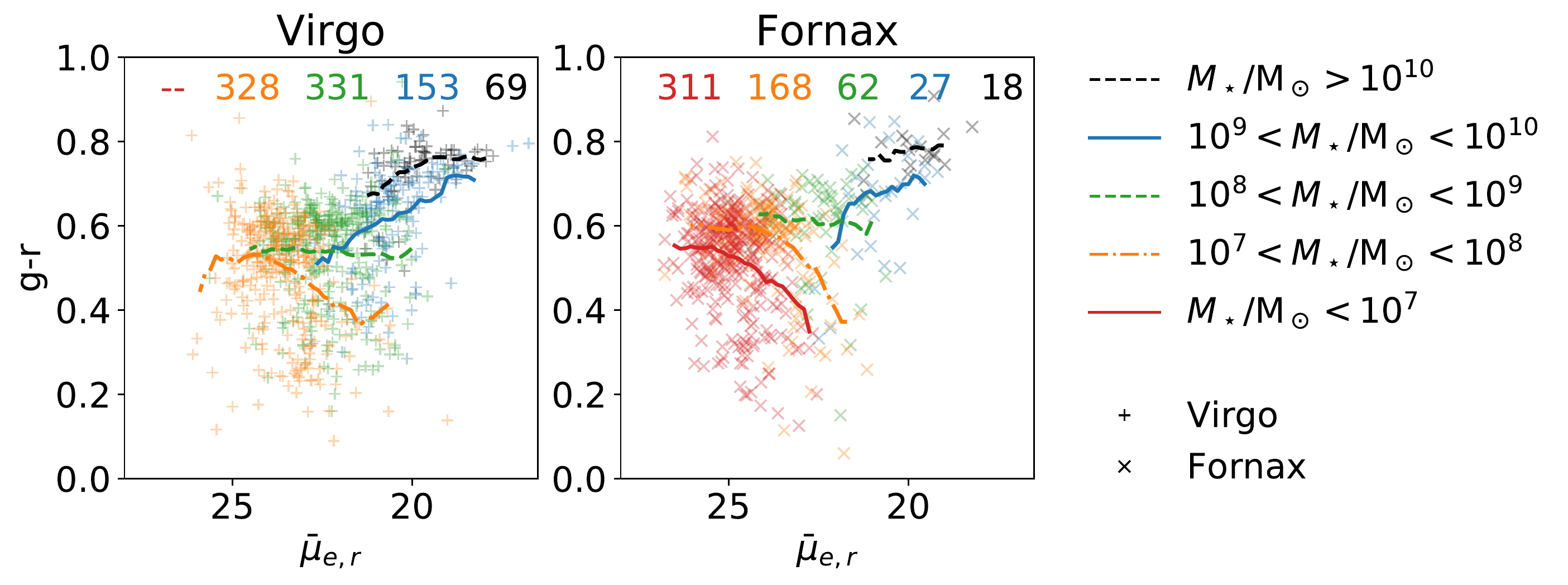}
   \caption{Mean effective surface brightness versus $g-r$ colour for cluster galaxies separated into stellar mass bins,
   for the Virgo \emph{(left panel)} and Fornax \emph{(right panel)} galaxy clusters.
   The numbers above the points quantify the sizes of the sub-samples.
   \mods{The curves display the running means of the relation in the  mass bins indicated on the \emph{right}, with 
   a window width of 1 mag arcsec$^{-2}$.
   The symbols, mass bins, and running means will also be used in subsequent
   figures.}
   Overall, the general trend of a change in the slope of the averaged data is found for both clusters.
   At low mass, the correlation of low surface brightness with redness was interpreted as 
   a result of fading and reddening after the star formation is quenched \citep{2019A&A...625A.143V}.\label{fig:vir_vs_fnx}}
\end{figure*}

\mods{A natural explanation is that the quenching of the dwarf galaxies
is caused by processes related to the cluster environment itself.} 
The prime example is the ram pressure on the inter-stellar medium as a galaxy moves  
through the hot cluster gas \citep{1972ApJ...176....1G}. When this ram pressure exceeds the
gravitational binding force that holds the gas in place, the gas is stripped and the
galaxy \modsTwo{loses} its reservoir of material for forming stars.
\mods{While there are other processes that can lead to gas loss (e.g.~tidal stripping of gas and feedback from supernova winds; \citealt{2010MNRAS.402.1599S}\footnote{\mods{We note that, even for the low-mass Local Group dwarf spheroidals, doubts have been raised regarding whether or not the supernova feedback suffices to prohibit star formation, e.g.~in \citet{2013MNRAS.433.3539G}.}}), the loss of interstellar gas due to the interaction with the intra-cluster gas will inevitably happen in galaxy clusters, \modsTwo{it} is specific to the environment, and it can be expected to dominate in galaxy clusters over the  other processes \citep[e.g.][]{2006ApJ...647..910H}. It can also be expected to have effects in even the  far outskirts of galaxy clusters \citep{2013MNRAS.430.3017B}.
Indeed, it has been suggested that the population of early-type \mods{dwarf} galaxies
may have formed by the transformation of recently fallen-in LTGs through this process \citep{2008ApJ...674..742B,2010ApJ...724L.171D}.}

\begin{table*}
\caption{Comparison of Virgo and Fornax.\label{tab:cluster}}
\begin{center}
\begin{tabular}{lccccc}
Cluster & Mass & Velocity dispersion & Relative gas density & Massive galaxies & Central galaxy density\\
\hline
Virgo & $1.7 \times 10^{14}\, {\rm M}_\odot$ & 753 (1340) km/s & 1 & 69 & 250 Mpc$^{-3}$\\
Fornax & $6 \times 10^{13}\, {\rm M}_\odot$ & $356\pm31$ ($405\pm45$) km/s& 1/4 & 18 & 500 Mpc$^{-3}$\\
~
\end{tabular}
\label{table}
~

\begin{minipage}{0.95\textwidth}\footnotesize{{\bf Notes:} The mass estimates are the fiducial values in \citet[see also references therein]{2011MNRAS.416.1197W}.  The velocity dispersions are given for the ETGs and, in parentheses, for the LTGs \citep{2006PASP..118..517B,2001ApJ...548L.139D}. The relative gas density is based on the logarithmic gas density profiles in \citet[their Fig.~6]{1997ApJ...482..143J}, which are approximately offset by a constant factor as a function of radius in linear physical units. The number of massive galaxies is that of the galaxies with $M_\star >10^{10} \, { \rm M}_\odot$ in our sample. The central galaxy number densities are from \citet{1989Ap&SS.157..227F}.} \end{minipage}
\end{center}\end{table*}%

The effects of ram pressure stripping are clearly observed   
for galaxies that still have gas, for example directly by the distorted  
gas distribution and stripped tails of gas. It can also be observed more indirectly by quantifying 
a deficiency of gas with respect to typical \mods{(late-type)} disk galaxies of the same stellar mass; this deficiency correlates
well with the exposure of a galaxy to the hot cluster gas \mods{\citep[e.g.][]{2004AJ....127...90Y,2007ApJ...659L.115C,2009AJ....138.1741C,2013AJ....146..124H,2018MNRAS.476.4753J}.
This even applies to the more tightly bound molecular gas \citep[e.g.][]{2019MNRAS.483.2251Z}.
It is much harder to \modsTwo{observe signatures of} this process in early-type (dwarf) galaxies (with the exception
of when they have recently accreted gas; \citealt{2017ApJ...840L...7S}),} which,
in this scenario, have\modsTwo{ already} been transformed and lost their gas, possibly long ago.

The strength of the ram pressure \citep{1972ApJ...176....1G}
depends on the cluster gas density  ($\rho_{\rm ICM}$) and the velocity of the galaxy through the cluster ($v$). \modsL{The  stripping of the gas from a galaxy occurs when the ram pressure exceeds the gravitational forces, that is when the following ram pressure criterion is met:
\begin{equation}
p_{\rm ram} = \rho_{\rm ICM} v^2 > 2\pi\,G\,\Sigma_\star \Sigma_{\rm ISM},
\end{equation} 
with the stellar and inter-stellar medium (ISM) surface densities $\Sigma_\star$ an $\Sigma_{\rm ISM}$.}
Simulations have repeatedly shown this analytic expression to be a good approximation (e.g.~\citealt{2009A&A...502..427V}, and most recently \citealt{10.1093/mnras/staa775}). Both the cluster gas density and the velocity dispersion of the galaxies scale with the mass of  
the galaxy cluster. This means that the comparison of the galaxy populations
in different galaxy clusters can potentially constrain the role of ram 
pressure in the shaping of these populations \citep[see also e.g.][]{2003ApJ...591...53T}.

The Virgo and Fornax galaxy clusters are the closest galaxy clusters to
us, at distances of 16.5 and 20.0 Mpc, respectively \citep{2005ApJS..156..113M,2009ApJ...694..556B}.
Both clusters have been well studied, with their (brighter) member galaxies
catalogued in the 1980s (Virgo Cluster Catalog -- VCC, \citealt{1985AJ.....90.1681B}; and Fornax Cluster
Catalog -- FCC, \citealt{1989AJ.....98..367F}) and some observed with the \emph{Hubble space telescope}  in the 2000s \mods{\citep{2004ApJS..153..223C,2007ApJS..169..213J}.} These clusters were recently surveyed with wide field-of-view cameras, resulting in new deep multi-band images
\mods{(e.g. \citealt{2012ApJS..200....4F} for Virgo, \citealt{2015ApJ...813L..15M}; \citealt{2016ApJ...820...42I,2018A&A...620A.165V} for Fornax) }that map out cluster galaxies down to low stellar masses and surface brightnesses.
\mods{\citet{2019A&A...625A.143V}} found evidence in the photometry
of the low-mass galaxies for ram pressure stripping in their past.
In this paper, we revisit their analysis, apply it to the
Virgo cluster, and compare the two clusters in terms of their galaxies' correlations between colour and surface brightness.

%--------------------------------------------------------------------
\section{Premise of the analysis}
\mods{
In this paper, we compare the (dwarf) galaxy populations of the Virgo and Fornax clusters.
These two \modsTwo{clusters} differ in mass by approximately a factor of five. The strength of the ram pressure
scales with the density of the intra-cluster medium (ICM) and the velocity of the galaxy squared (equation 1).
The typical random velocities of
galaxies in Virgo can be expected to be more than \modsTwo{twice} as large as in Fornax \citep[e.g.][]{2007ApJS..169..213J}. 
Together with the denser ICM, which is denser by a factor of $\sim$4  \citep{1997ApJ...482..143J,1999A&A...343..420S,2002ApJ...565..883P}, 
the ram pressure can be expected to be more than an order of magnitude stronger for the Virgo cluster (see Table \ref{tab:cluster}; \citealt{2013MNRAS.428..834D} estimated a factor of 16). 
On the other hand, for harassment by galaxy encounters and the cluster tidal potential, the difference in strength is expected to be much smaller.\ In fact, it is less clear in which cluster \modsTwo{these processes can be expected to be more efficient} since, despite the difference in mass, both reach similar central  mass densities
 \citep{2007ApJS..169..213J,2008MNRAS.388.1062C} and Fornax's central galaxy number density  even
 exceeds that of Virgo \citep{1989Ap&SS.157..227F}.}
 
\modsL{ The first goal of this paper is to  search for differences in the dwarf galaxy populations of the two clusters. 
Thanks to the differences  between the properties between the Virgo and Fornax clusters, these could the be used to
 measure  the different strengths of transformative processes \mods{that act in cluster environments.}}
 
\mods{The other goal concerns the more comprehensive analysis of the actual evidence 
for the passive evolution of low-mass galaxies that was suggested by \citet{2019A&A...625A.143V}.
They found that when the galaxies are binned by brightness or stellar mass in plots of galaxy colour versus surface brightness, the galaxies show a varying behaviour depending on the mass bin.
While galaxies with high masses show a correlation of high surface brightness and redness, 
this trend reverses for the low-mass galaxies. For these, a redder colour is accompanied by a fainter surface brightness.
\citeauthor{2019A&A...625A.143V} interpreted this behaviour at low galaxy masses as a consequence 
of the quenching by the cluster environment 
and subsequent fading and reddening.
Their plot for the Fornax cluster is reproduced (with updated data and aperture colours; \citealt{su2021fornax}, Salo et al. in prep.) and extended to higher masses in Fig.~\ref{fig:vir_vs_fnx} (right panel). The galaxies in bins with $M_\star<10^8\, {\rm M}_\odot$ clearly show the correlation between redness and faint surface brightness, while the more massive galaxies do not.
}

\begin{figure}
  \centering
  \includegraphics[width=0.45\textwidth]{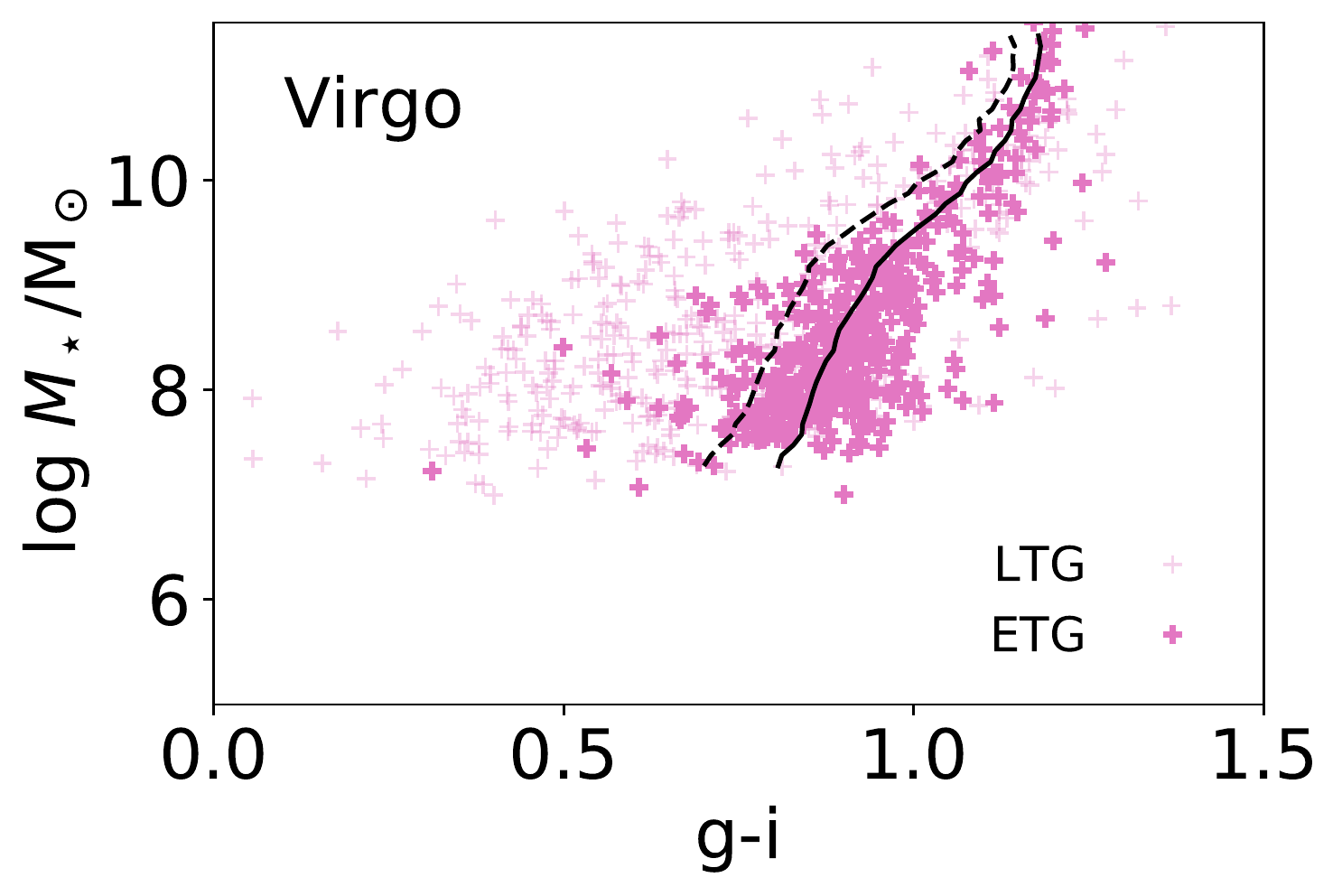}\\
  \includegraphics[width=0.45\textwidth]{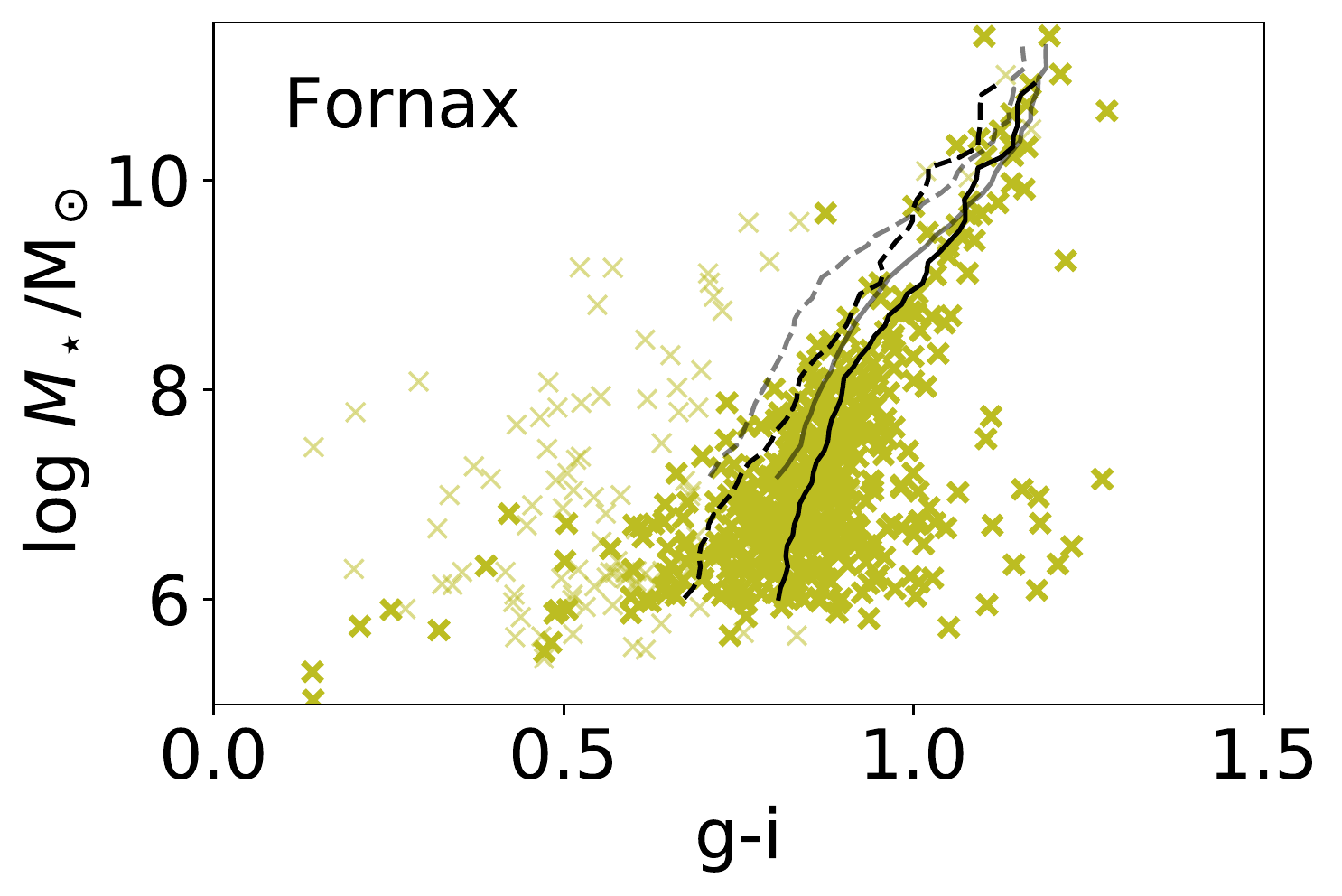}
   \caption{Stellar mass versus $g-i$ colour relations for galaxies in the Virgo \emph{(top)} and Fornax \emph{(bottom)} clusters (for other colours, see Fig.~\ref{fig:cmrall}). The solid black lines show the running mean relations \mods{(in $M_\star$ with a bin width of 1 dex)} for the ETGs \emph{(bold symbols)}, while the dashed lines
   indicate the mean minus one $\sigma$. The Virgo relations are shown (in grey) in the lower panel for reference. 
\mods{The relations are largely consistent for the Fornax and Virgo clusters.
}\label{fig:cmr}
    }
\end{figure}

\section{Data}
\mods{Our  analysis  is based on data and catalogues available in the literature, stemming from two different data sets. First, for the Virgo cluster, the photometry and other basic parameters are collected from several studies that made
use of Sloan Digital Sky Survey (SDSS) data, in particular its fifth data release \citep{2007ApJS..172..634A}:
The data for the early-type galaxies (ETGs) originate from \citet{2008ApJ...689L..25J} and \citet{2009ApJ...696L.102J}, and the source for the LTGs is \citet{2014A&A...562A..49M}.\footnote{\modsL{Data from the Next Generation Virgo Cluster Survey (NGVS; \citealt{2012ApJS..200....4F}) with improved depth exist,} and the photometric parameters for the central part of the cluster recently became  publicly available \citep{2020ApJ...890..128F}. The overall consistency of the SDSS- and NGVS-based parameters was documented by e.g.~\citet{2017ApJ...836..120R}. \modsL{For} the purpose of this study, the combination of the SDSS sample for Virgo and the more recent, deeper Fornax sample, as described, turns out to be useful (the statistics are improved by the many bright galaxies in Virgo and the large number of lower-mass systems in the deeper Fornax data; the total number of galaxies in the sample for Virgo exceeds that for Fornax by a factor of approximately two).} 
 We refer the reader to these studies as well as \citet{2007ApJ...660.1186L} for details on the data reduction and measurements, but we list a few salient characteristics in the following. The root mean square (rms) noise levels per pixel ($0\farcs396^2$) converted to surface brightnesses correspond to 24.2, 24.7, 24.4, 23.9, and 22.4 mag arcsec$^{-2}$ in the five SDSS filters $u$, $g$, $r$, $i$, and $z$, respectively. While the SDSS data are not particularly deep by today's standards, due to the relative short  exposure time of effectively 54s with a moderately sized telescope, its background is very uniform thanks to the observing strategy. The median full-width at half-maximum (FWHM) of the seeing point-spread function is $1\farcs$4 in the $r$-band. The colours are determined in elliptical apertures of two effective radii  \citep{2009ApJ...696L.102J} and one Petrosian semi-major axis \citep{2014A&A...562A..49M}. These two apertures roughly coincide for exponential surface brightness profiles. The morphological classifications are taken from the original VCC (visual classifications on photographic plates in $B$), with updates for ambiguous cases from the aforementioned studies (based on multi-wavelength SDSS images). Classifications of any sub-classes of E, S0, dE, and dS0 are here considered as early types, while any galaxy with a classification from Sa to Im, including blue compact dwarfs (BCDs), is considered as a late type. \citet{2014ApJS..215...22K} visually re-classified the morphology of Virgo galaxies based on SDSS multi-band images. In terms of our broad late- versus early-type classification, fewer than 2.5\% of cases (24 excluding uncertain classifications) differed between the VCC and the updated classifications (see also examples in \citeauthor{2014ApJS..215...22K}), several of which were already noted in the studies cited above.
 \citet{2011MNRAS.416.1197W} discuss the completeness of the used Virgo data in detail and conservatively estimate that the data are complete at least to a brightness of $M_r=-15.2$ mag in the $r$-band. This was confirmed with the deeper data of \citet{2012A&A...538A..69L}, who did not discover any galaxies brighter than $M_r<-13$ mag that are not already listed in the VCC.
Our assumed  distance to the Virgo cluster of 16.5 Mpc  corresponds to a distance modulus of $m-M=31.09$ mag \citep{2009ApJ...694..556B}.}

Second, for the Fornax cluster, the data were collected with OmegaCAM on ESO's  \modsL{Very Large Telescope Survey Telescope in the Fornax Deep Survey  \citep[FDS][]{2016ApJ...820...42I,2018A&A...620A.165V}.} The in-depth details for this data set can be found in \citet{2018A&A...620A.165V,2019A&A...625A.143V}, with minor updates in \citet{su2021fornax}. These images are substantially deeper than SDSS, with rms noise levels of 26.6, 26.7, 26.1, and 25.5  mag arcsec$^{-2}$ per $0\farcs2^2$ pixel in the $u$-, $g$-, $r$-, and $i$-bands, respectively. The typical seeing conditions led to average FWHMs of $1\farcs2$, $1\farcs1$, $1\farcs0$, and $1\farcs0$ in the various filters in the same order.
{For the bright galaxies, the classifications are from the original FCC and were visually checked by \citet{su2021fornax} with FDS data.}
\citet{2018A&A...620A.165V} visually classified the morphology of dwarf galaxies broadly into \emph{i)} smooth ETGs,  \modsTwo{\emph{ii)}} ETGs with structures (here subsumed into the early-type sample), and \emph{iii)} LTGs with blue features, star-forming clumps, and irregularities (see their Fig.~18 for sample images of galaxies in each class). Their visual classification  is based on $gri$-colour and residual (after the subtraction of a simple \modsTwo{S\'ersic} model) images. For a small number of uncertain cases (22), parameters such as the concentration, colour, and residual flux fraction (quantifying how well the image is represented by a simple smooth model) were considered for the final classification.
\citeauthor{2018A&A...620A.165V} also extensively tested the detection completeness and declared a 50\% completeness at $M_r = -10.5$ for galaxies larger than 2 arcsecs.
The colours were determined in elliptical half-light apertures  (Salo et al. in prep.). This difference in comparison to the Virgo cluster data  does not introduce any systematic effects into the analysis.  The distance of 20.0 Mpc  to the Fornax cluster corresponds to a distance modulus of $m-M=31.51$ mag  \citep{2009ApJ...694..556B}.

\mods{Benefitting from the available multi-band data, we converted the absolute magnitudes to
stellar masses using the conversion found in \citet{2011MNRAS.418.1587T},
\begin{equation}
\log M_\star/M_\odot  = 1.15 + 0.7 \times (g-i)- 0.4 \times M_i,
\end{equation}
with the $g-i$ colour and the total absolute $i$-band magnitude $M_i$. A representation of the data sets is given 
in Fig.~\ref{fig:cmr}.
The colour--mass relations are similar for both clusters (see also \citealt{2017ApJ...836..120R,2018A&A...620A.165V}).
 \mods{There is a small offset, which depends on the colour and galaxy mass (see also Fig.~\ref{fig:cmrall} in the appendix). While it would be tempting to speculate that the somewhat bluer colours (e.g.~$\sim0.05$ mag in g-i; see Fig.~\ref{fig:cmr}) of galaxies in the Virgo cluster are related to its supposedly more recent assembly (see discussion), the offset is not significant in terms of  calibration uncertainties and similar systematics. For the rest of the analysis, the offset will be ignored.}}

Throughout this text, we also use the mean effective surface brightness $\bar\mu_{e,r}$  in the $r$-band, which is defined as the average surface brightness within an elliptical aperture containing half of the total light.
All of the above studies from which we gathered the data used the $r$-band as a fiducial filter with the best combination of $S/N$ and seeing FWHM. We followed this lead and use the $r$-band whenever brightness or surface brightness is plotted throughout our analysis. For both clusters, the samples include galaxies from the entire cluster, ignoring known substructures of the Virgo and Fornax clusters. 
%--------------------------------------------------------------------

\begin{figure*}
  \centering
\includegraphics[width=\textwidth]{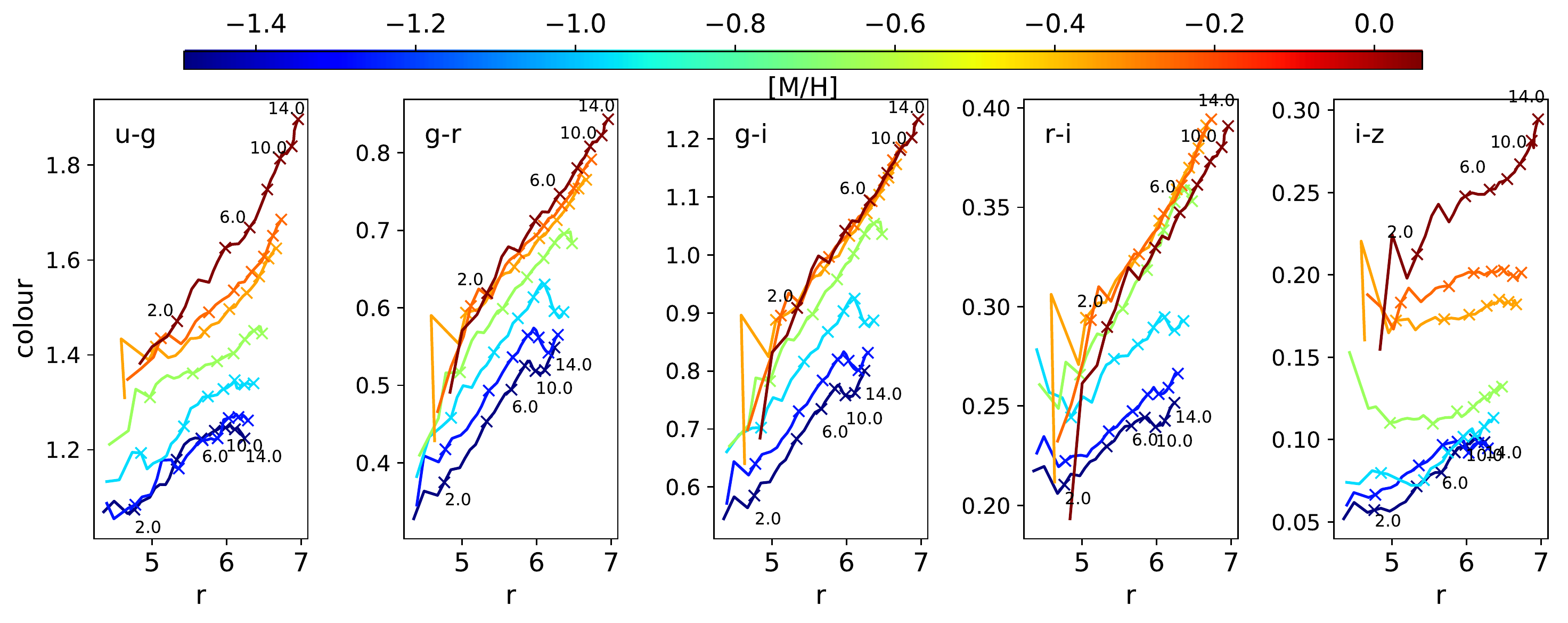}
   \caption{Predictions of how passively evolving single stellar population (SSPs)  fade and redden. The different panels show various colours based on SDSS filters (see the top left label in each panel) for time steps corresponding to ages older than 1 Gyr. \mods{The SSP models'  predictions (see text) are shown for a range of metallicities from $[M/H]=-1.5$ to solar (see colour bar on top; [M/H]$=0$ corresponds to a 2\% mass fraction of the metals), and} the ages are marked in steps of 2 Gyrs (see also the numbers in the panels above and below the model lines). \mods{In our analysis, we compare the slopes of the lines (fitted over an age range of 2 to 12 Gyrs; see the table in the appendix) to the observations. In each panel, the slopes are similar for different metallicities. It can also be seen that the absolute value of the colour is determined by the metallicity and age of the stellar population. The magnitudes are normalised \modsTwo{per unit solar mass of the initial stellar population.} For unchanged physical sizes of the galaxies, differences in magnitudes in the brightness correspond to differences in surface brightness. } \label{fig:stellpop}}
\end{figure*}

%--------------------------------------------------------------------
\section{Analysis}
%--------------------------------------------------------------------

\mods{
\citet{2019A&A...625A.143V} observed a correlation between redness and decreasing surface brightness for 
dwarf galaxies and interpreted this as an effect of passively evolving stellar populations whose star formation has already been quenched. 
 In order to scrutinise the  interpretation, 
\modsTwo{here  we} want to analyse colours from the whole wavelength
coverage of the available multi-band data and compare them to  predictions of  stellar evolution
models \citep[][retrieved from \url{http://miles.iac.es}]{2010MNRAS.404.1639V,2012MNRAS.424..157V,2012MNRAS.424..172R,2016MNRAS.463.3409V}. The stellar evolution calculations are based on BaSTI isochrones \citep[Bag of Stellar Tracks and Isochrones,][]{2004ApJ...612..168P}, solar $\alpha$-element abundances, and a revised Kroupa initial mass function \citep{2001MNRAS.322..231K}.
\mods{The models output the brightnesses of the stellar population in different filters at various evolutionary stages, which are used to calculate the colours.}
Figure~\ref{fig:stellpop} shows the fading and reddening tracks  of stellar populations of different metallicities (from $[M/H]$$\sim-1.5$ to about solar) in various SDSS colours. The  tracks mostly show a relatively consistent slope -- a fading reddening track slope (FRTS) -- which we will use for comparisons
to the data. The slope of each track is parametrised by fitting a linear relation in the age
range from 2 to 12 Gyrs.
}

\mods{We emphasise that the intention here is neither to model the quenching of star formation itself, nor to draw any conclusions as  to what caused the quenching. Instead, the aim is  to analyse how changes in colour and (surface) brightness (i.e.~the reddening and fading) 
are related in the picture of a passively evolving stellar population and to test whether or not this is compatible with the observed correlations. \modsTwo{Colour depends both on age and metallicity. However, as shown in Fig.~\ref{fig:stellpop},} the correlations between colour and brightness (i.e. the slopes or FRTS) are largely independent of the metallicity. Instead of analysing the absolute colour and surface brightness, we focus on their correlations, which are compared to the FRTS. It should also be noted that, while in Fig.~\ref{fig:stellpop} the colours are plotted as a function of $r$-band brightness (normalised to a stellar population of an initial mass of $1{\rm M}_\odot$), any difference in this brightness translates to an equal difference in surface brightness for a stellar population of a galaxy as long as the area over which it is spread remains unchanged.  }

\begin{figure*}
  \centering
\includegraphics[width=0.75\textwidth]{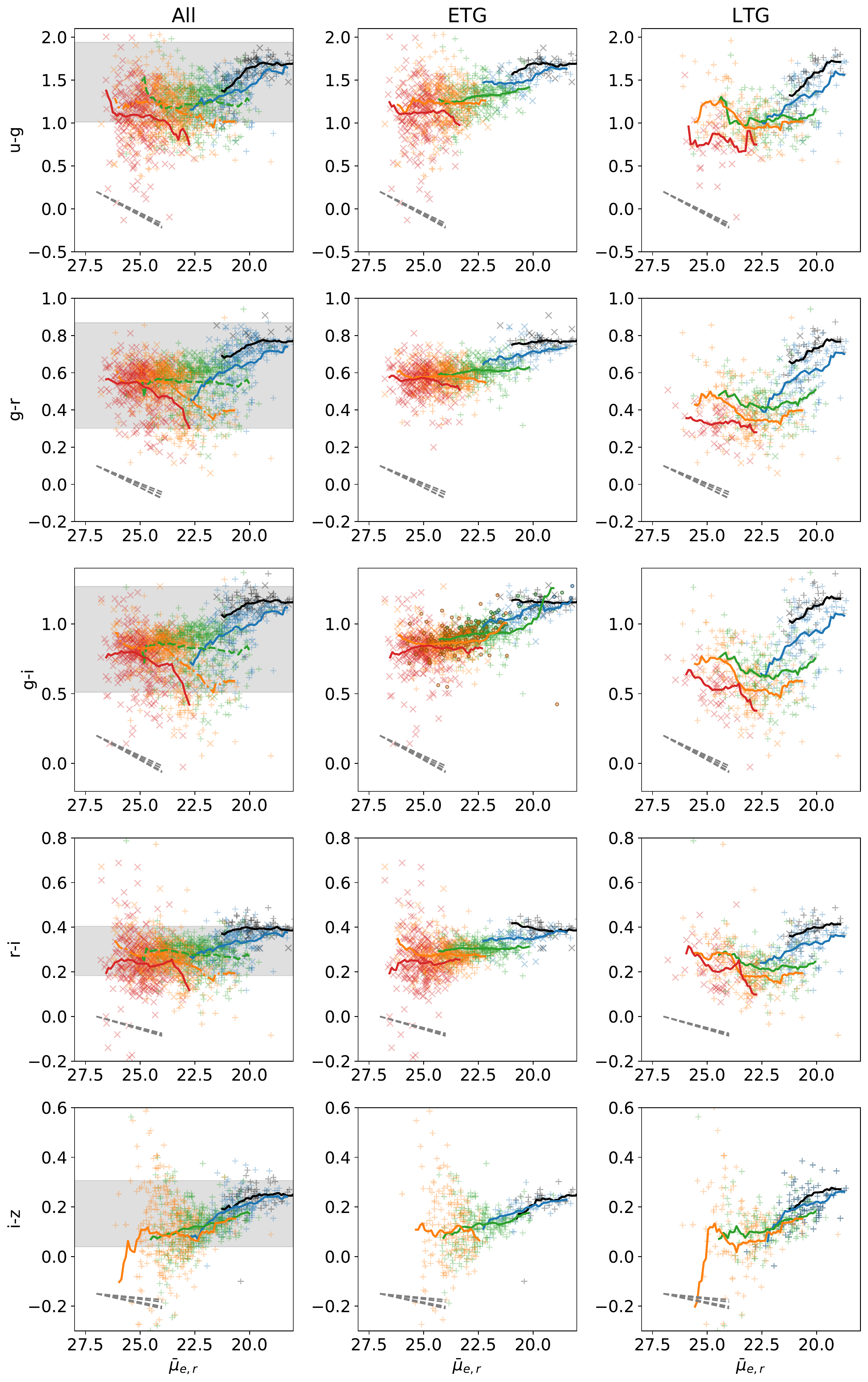}

   \caption{Colour versus mean effective surface brightness; symbols and mass bins are the same
   as in Fig.~\ref{fig:vir_vs_fnx}. The galaxies for the Virgo and Fornax
   clusters are jointly analysed \emph{(left column)}, but ETGs (\emph{middle}) and LTGs (\emph{right}) are separated (symbols and colours as in Fig.~\ref{fig:vir_vs_fnx}; in the ETG panel for $g-i$, \mods{data for the Perseus cluster are added and displayed as circles; for a detailed view, see Fig.~\ref{fig:perseus} in the appendix)}. Different \mods{galaxy} colours are shown in the various rows.
   In addition, the slopes of passively fading and reddening SSP models for a range of metallicities are indicated in the bottom left of each panel  (see text and Fig.~\ref{fig:stellpop}; \mods{for better readability, they are plotted at a position offset from the data)}; they are very similar for all metallicities in each panel. The y-ranges from Fig.~\ref{fig:stellpop} are shaded in grey in
  the left panels for reference.
  \mods{While the characteristic change of the average slope across the mass bins can be seen for the LTGs, the ETGs show only a mild (or no) variation.}
  The slope for the low-mass LTGs is largely consistent with the SSP models.\label{fig:allSB}
  }
\end{figure*} 

\subsection{Comparison of Virgo and Fornax in the light of the FRTS}
In the left panel of Fig.~\ref{fig:vir_vs_fnx}, the galaxies of the Virgo cluster are analysed in the same way as the Fornax galaxies in  \citet{2019A&A...625A.143V}: They are binned by their stellar masses and, \mods{for each bin,} their colours are plotted versus their surface brightness.
The overall behaviour closely matches that of the  galaxies in the Fornax cluster. The average slope changes from the low-mass 
bins to the high-mass bins, so that,
for the lowest masses \mods{($<10^8\, {\rm M}_\odot$),} a brighter surface brightness correlates with a bluer colour, while for the higher-mass bins it correlates with a redder colour.
Interestingly, the highest-mass bin, which was not probed by \citet{2019A&A...625A.143V}, appears to flatten out to a constant red colour.

As a next step, we carried out \modsTwo{similar analyses} for colours covering the full spectral range of the multi-band SDSS/FDS data. 
Furthermore, after \mods{establishing} the similarity between the Virgo and Fornax clusters in Fig.~\ref{fig:vir_vs_fnx}, we jointly analysed the data  from both clusters, which has the benefit of higher numbers of galaxies per bin and will allow for a further division of the sample into ETGs and LTGs as the next step. \mods{We note that  differences between the clusters could in principle go unnoticed in the mixed samples, which are dominated by the larger numbers of Virgo galaxies in the more massive bins and the Fornax galaxies in the lowest-mass bin. However, in later plots, the samples are also considered separately again.}
 The results for the various colours are shown in the left panels of Fig.~\ref{fig:allSB}. 
 \modsTwo{Particularly in the $g-i$ and $r-i$ colours (the latter being independent of $g-r$), but also in $u-g$, the characteristic behaviour of Fig.~\ref{fig:vir_vs_fnx} is seen. For $i-z$, this conclusion is more tentative,  likely due to the poorer data quality in the $z$-band,} for a $S/N$ that is a factor of ten worse for the Virgo dwarfs \citep{2007ApJ...660.1186L}, and the lack of Fornax data in $z$.
  \modsTwo{The slope of correlations} of colour and surface brightness in the low-mass bins are largely consistent with the FRTSs, which do not depend strongly on metallicity.
 The weaker trend of redder colours at fainter surface brightnesses in the $i-z$ colour as compared to those at shorter wavelengths 
 can be expected in this context \mods{since the FRTS of the single stellar populations (SSPs) is shallower as well.}

\begin{figure}
\centering
\includegraphics[width=0.48\textwidth]{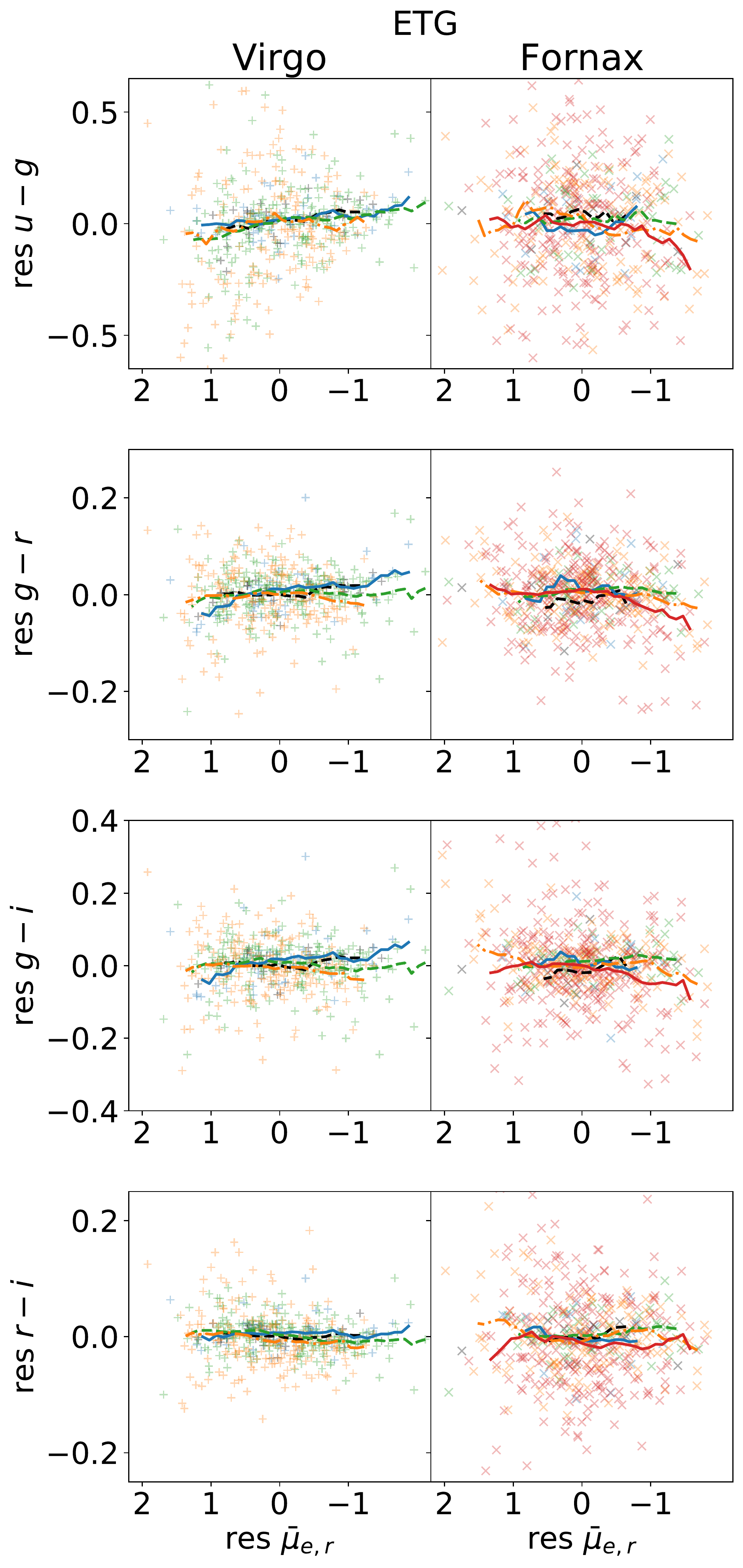}
   \caption{Residual colour versus residual surface brightness for
   ETGs in the Virgo \emph{(left)} and Fornax \emph{(right)}
   clusters. The residual quantities are obtained by subtracting  the mean
   trends between galaxy mass and the the colours and surface brightnesses (see text).
   The y-range is adjusted to half of that in Fig.~\ref{fig:stellpop}, with
   a few points for the lowest-mass bin falling out of the plotting range.
   The flatness of the relation for all bins (with the possible exception of \modsL{ the
   $u-r$ colour}) indicates that the behaviour of the ETGs in Fig.~\ref{fig:allSB}
   is determined by the scaling relations of colour and surface brightness
   with stellar mass.     Symbols and colours are the same as in Fig.~\ref{fig:vir_vs_fnx}.  \label{fig:ETGres}    }
\end{figure}

\subsection{Early-type galaxies}

The division into ETG and LTG sub-samples is shown in the middle and right columns of Fig.~\ref{fig:allSB}.
We first focus on the ETGs. The slopes for the different mass bins of ETGs resemble one another more closely, without the distinct change of slope from high to low galaxy mass seen in the left column. In fact, the slopes are relatively flat, close to zero.

Colour and surface brightness, which are plotted in Fig.~\ref{fig:allSB}, both  follow their own scaling relations with stellar mass. 
When plotting colour versus surface brightness binned over stellar mass, these scaling relations are imprinted in the behaviour of the galaxies in plots, such as those in Fig.~\ref{fig:allSB}. In order to remove the mass-related trends, we determined the mean scaling relations (per cluster and for each colour and \modsTwo{surface brightness}) by a running mean with respect to stellar mass (see e.g. Figs.~\ref{fig:cmr} \mods{and~\ref{fig:SB}}) and subtracted it off the colour and surface brightness values in order to analyse the behaviour of residual colour versus residual surface brightness.
Figure~\ref{fig:ETGres} shows that the slopes in the different mass bins for the ETGs all become flat.
In other words, the whole behaviour of ETGs in Fig.~\ref{fig:allSB} is explained by the colour and surface
brightness scaling relations with stellar mass without any indications of fading or reddening.
Returning to the highest-mass bin, which was unexplored by \citet{2019A&A...625A.143V},
the consistency and flatness of the residual relations for all mass bins also indicate that the flattening  in this highest mass \mods{in Fig.~\ref{fig:vir_vs_fnx}},
which is dominated by the ETGs, is real \mods{and} simply a consequence of the shape of the involved scaling relations.

\subsection{Late-type galaxies}
The curves for LTGs form a U-shape \mods{in Fig.~\ref{fig:allSB}, similar to} the characteristic behaviour seen in the left column and in Fig.~\ref{fig:vir_vs_fnx}. The slopes of the curves for the two lowest-mass bins appear to be consistent overall with the fading tracks from the stellar population models in all panels for the LTGs.

\begin{figure}
    \centering
  \includegraphics[width=0.48\textwidth]{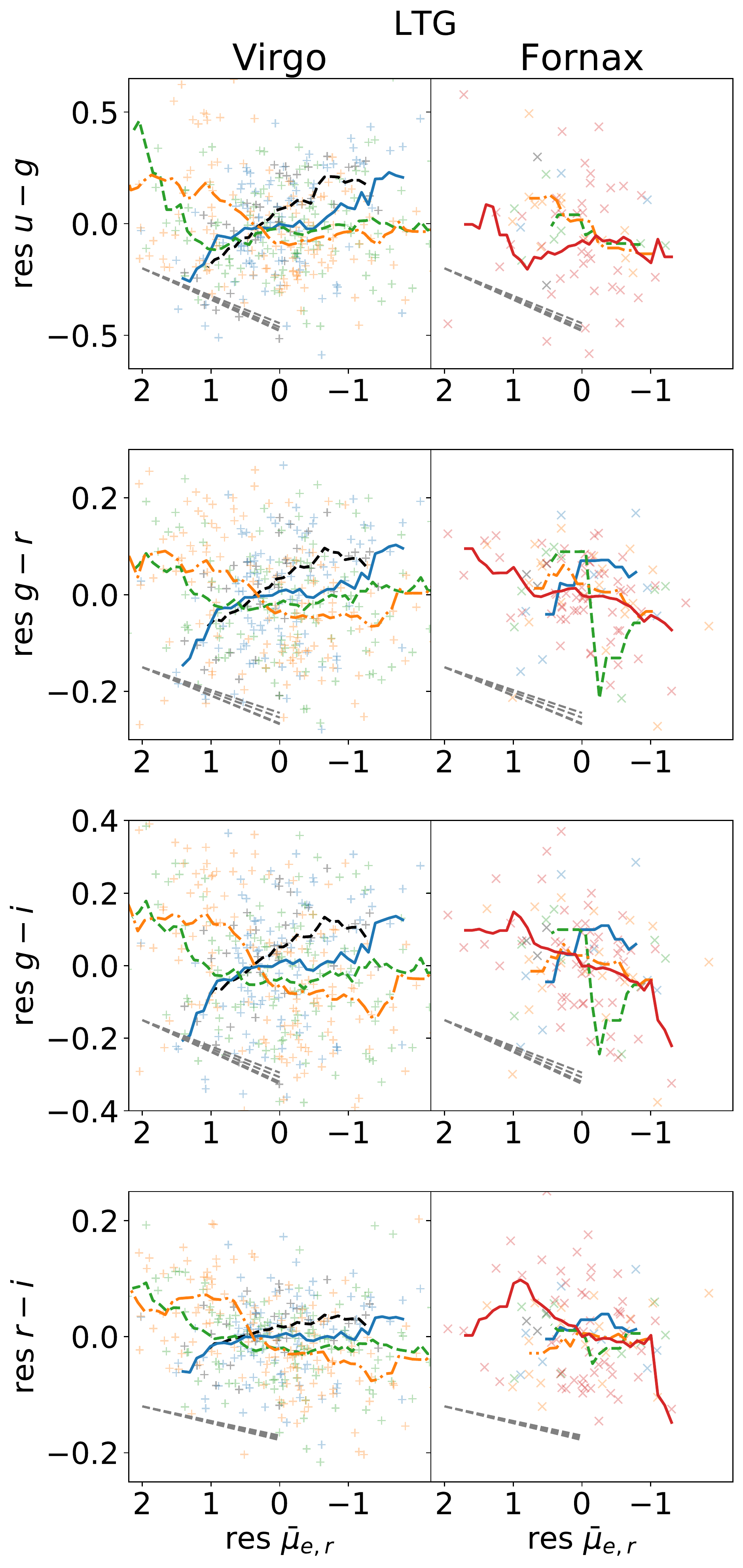}
     \caption{\mods{Residual colour versus residual surface brightness for LTGs (similar to Fig.~\ref{fig:ETGres} for ETGs).} The fading and reddening directions are repeated from Fig.~\ref{fig:allSB}. Symbols and colours are the same as in Fig.~\ref{fig:vir_vs_fnx}. \mods{Unlike the ETGs, the low-mass LTGs ($M_\star < 10^8\, {\rm M_\odot}$) also show slopes in the residual relations that are consistent with the FRTS.}\label{fig:LTGres}}
  \end{figure}

The smaller number of LTGs hampers their residual relation analysis
 but qualitatively confirms our findings. 
\mods{The residual curves for these galaxies in Fig.~\ref{fig:LTGres}, after the subtraction of the mean scaling relations of colour and surface brightness, confirm that the behaviour of the late types is not simply a consequence of the scaling relations as for the ETGs (see Fig.~\ref{fig:ETGres}). Instead, the low-mass LTGs are consistent with the FRTS. This supports the interpretation for the  low-mass LTGs with ageing stellar populations.
}

 As shown in the appendix, the corresponding analysis for surface mass densities instead of surface
 brightness (Fig.~\ref{fig:alldens}) also confirms the consistency of low-mass LTGs with this scenario. 
 The late types in the \mods{two} lowest-mass bins \mods{($M_\star < 10^8\, {\rm M_\odot}$)} 
 show a constant mean colour as a function of  mean effective surface mass density,
as expected, since the mass density is unaffected by the fading. \mods{We note that the non-zero slopes of
the average trends for these mass bins in the $u-g$ and $z-i$  are not significant given the large uncertainties
in $u$ and $z$.}
Interestingly, for the LTGs in the higher-mass bins, it seems possible that the relations in all bins turn 
redwards \mods{above a  surface mass density  of around 50-100\, ${\rm M} _\odot {pc}^{-2}$}
 for galaxies in the mass bins that reach these densities \mods{(e.g. the bins with $M_\star > 10^8\, {\rm M_\odot}$).}

%-------------------------------------------------------------------
\section{Discussion}
Our first point was that the galaxies in the Fornax and Virgo clusters behave very similarly in
Fig.~\ref{fig:vir_vs_fnx}. The two clusters differ in mass and other properties \mods{(Table \ref{tab:cluster}),} which means
that the physical processes (e.g.~ram pressure stripping and harassment) that may let the clusters drive the evolution of the galaxies
can be expected to have different strengths and different relative weights when \mods{the clusters are} compared to
each other. 
This makes the similarity of the two clusters in Fig.~\ref{fig:vir_vs_fnx} surprising at first glance.
However,\mods{ when interpreted} as the fading and
reddening \modsL{of the low-mass galaxies}, this similarity is a consequence of the \mods{passive} evolution of the stellar populations, \modsTwo{which, after quenching is completed, proceeds independently of the environment and the processes that quenched the star formation. In other words,} the galaxies that are redder and have fainter surface brightnesses were quenched earlier, no matter how they were quenched or how many of them were quenched (more about that below).
In fact, in the cosmological simulations of \citet{2020MNRAS.497.2786T}, \mods{there is also a correlation between central surface brightness and mean stellar age in the inner kiloparsec, even for galaxies in field environments.}

One aspect for which the differences \mods{between the clusters} (for example the strength of the possible ram pressure) can  be expected to matter
is the galaxy masses at which the LTGs are affected so as to make their colour versus brightness relation compatible with the FRTS (see also \citealt{2019A&A...625A.143V}). 
When analysing the two clusters separately, the LTG samples binned by stellar mass become relatively small, prohibiting any strong conclusion in this respect. Even so, it appears at least possible that the fading and reddening trend is  already followed for Virgo at higher galaxy masses compared to Fornax (see Fig.~\ref{fig:LTGres}, e.g.~the green lines for $10^8  <{\rm M}_\odot/M_\star < 10^9 $), which, if true, could be interpreted as an imprint of the stronger ram pressure stripping in the Virgo cluster. 

In order to more quantitatively verify the viability of the fading and reddening interpretation for the low-mass (late-type)
galaxies, we compared the galaxies' average behaviour in Fig.~\ref{fig:allSB} to the FRTS predicted for the passive evolution
of quenched (single) stellar populations and found a \modsTwo{reasonable} agreement for the low-mass LTGs.
On the other hand, the early-type galaxies do not show the same behaviour\footnote{\modsTwo{We note that this statement is less clear for the longest wavelength colour $i-z$, for which the expected slope is shallowest and  the data in terms of sample size, mass limit, and data quality are the worst.}} despite the fact that the stellar population evolution
model suggests that they should.
This applies even if the quenching event had occurred in the more distant past (with the possible exception of a very early quenching, i.e. $>10$ Gyr ago).
However, the use of SSP models here could be an over-simplification (e.g.~Local Group dwarf galaxies 
have been shown to often have long-lasting star formation \mods{with sporadic episodes of enhanced activity}; \citealt{2014ApJ...789..147W}). To test this, we qualitatively explored what effect more complex 
star formation histories \mods{can have in Appendix \ref{app:csp}. The FRTS is largely consistent with that of the SSP models, and the details of the star formation history prior to the quenching have a relatively minor impact on our analysis. }

For the ETGs, we found  that their behaviour  in Fig.~\ref{fig:allSB} is fully determined by their 
scaling relations of colour and surface brightness with stellar mass. 
This appears to apply generally since it was also the case for a third galaxy cluster that we looked at \mods{(the Perseus cluster, which is more massive than Fornax and Virgo; see Fig.~\ref{fig:additional}).}
\mods{This means that the early-type dwarfs follow a colour-magnitude relation that is driven by a mass-metallicity relation (and possibly changes of mean stellar age as a function of mass) and a mean surface brightness-mass relation, but that the offsets from these two relations are uncorrelated. This is different for the LTGs, for which the
 same analysis} shows that the FRTS is also clearly visible in the residuals  (Fig.~\ref{fig:LTGres}). 

It is quite puzzling \mods{why the colour--surface brightness relation of the ETGs does not show any resemblance to the FRTS, regardless of whether or not they have recently  been transformed from LTGs, given the fact that they predominantly have passively evolving stellar populations.}
First, it needs to be kept in mind that the SSP models are in some cases not well approximated by 
our linear fits for the oldest ages; as such, one possible explanation could include a very early quenching.
Other possibilities include a washing out of any such trends by other processes that redistribute the stars, 
and which therefore can modify the \mods{stellar surface density} independently of the colour.

An early transformation to quiescence may be consistent with the conclusions reached by
 \citet{2013MNRAS.432.1162L}  by comparing the cluster populations to sub-halos in the 
 Millennium-II $N$-body simulations \citep{2009MNRAS.398.1150B}.
 \mods{\citeauthor{2013MNRAS.432.1162L}  identified dwarf early-type, transition-type, and late-type galaxies with
 sub-halos that became part of a massive halo early on, at intermediate
 ages, or at more recent times. \modsTwo{The} limits were chosen to reproduce the observed fractions
 of various classes. The projected spatial distributions of the three dwarf galaxy populations with respect
 to the cluster centre and the projected spatial distributions of the three populations of sub-halos were found
 to \modsTwo{be qualitatively compatible (with early types \modsL{and  early infall sub-halos}  being more centrally concentrated; this agreement was only found} for the identification in the order described above).
 \modsTwo{We note that the abovementioned criterion to classify the sub-halos in the simulation specifically took into account
 the time it took for the halo to became a sub-halo of a massive halo, the latter not necessarily being  the parent halo of the final snapshot.
For the observed galaxies, this serves as a reminder that they may also have suffered pre-processing in galaxy groups that then
 fell into the Virgo or Fornax clusters.
 The consistent radial distributions of sub-halos \modsL{and} dwarf types   were similarly 
 obtained by \citet{2013MNRAS.432.1162L} with an alternative identification of the dwarf sub-classes with
 sub-halos that experienced significant ($\rightarrow$ early types), marginal ($\rightarrow$ transition types), and } no significant dark matter mass losses ($\rightarrow$ late types).}
 This suggests that both factors, an early quenching and processes that  change the stellar surface 
 densities, can be expected to be at work at the same time: \mods{One process that can alter the 
 stellar surface densities without necessarily having an effect on the stellar populations  
 is harassment, which is only expected to significantly affect the stellar body of a galaxy when 
 significant amounts of the dark matter are stripped \citep[e.g.][]{2015MNRAS.454.2502S}.}

 With regards to the problem of early-type dwarfs not showing any resemblance to the FRTS, \mods{the two abovementioned factors} work hand in hand:  Even for ages for which the linear fits of the stellar evolution tracks are
 a good representation, the amount or rate of the evolution along the track becomes increasingly smaller \modsL{and} slower the further in the past the quenching event occurred. 
For instance, the age span from 2 to 6 Gyrs stretches approximately twice the distance along the track than that from 6 to 10 Gyrs.\footnote{
 This argument also applies in a similar way when \modsTwo{considering models with extended star formation histories (see Appendix \ref{app:csp}).}} 
 In other words, a weaker trend can more easily be washed out, and other processes (i.e.~those that can alter the surface densities, such as harassment) can 
 more easily dominate the evolution in the colour--surface brightness space.

\begin{figure}
\centering
\includegraphics[width=0.48\textwidth]{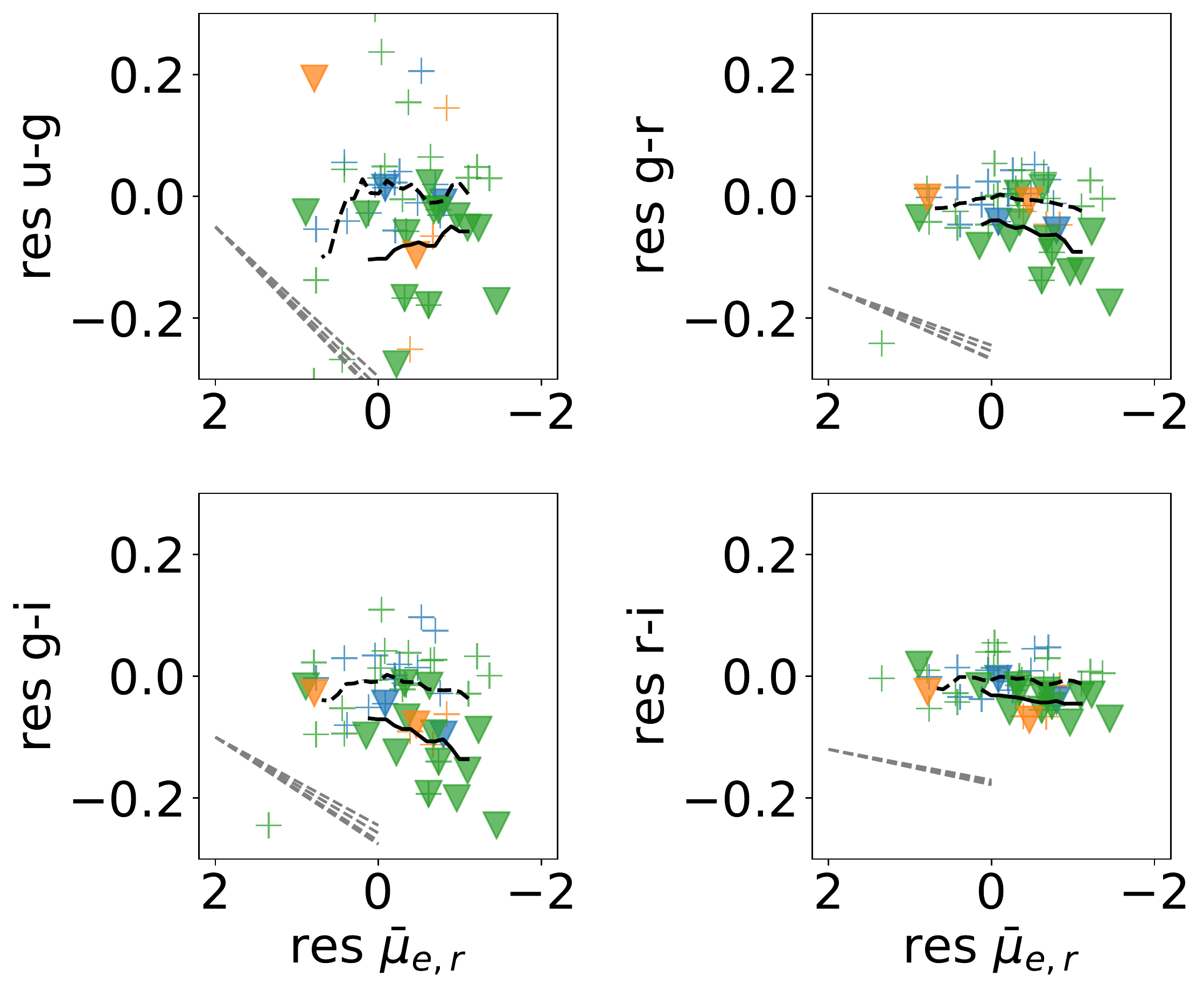}
   \caption{\mods{Residual colour versus residual surface brightness for
   ETGs in the Virgo cluster with disk features \emph{(plusses)} and blue cores \emph{(triangles)},
   i.e. the plots in the left panels of Fig.~\ref{fig:ETGres} for this sub-sample of galaxies.}
   The \emph{ dashed} and \emph{solid black} \mods{curves} show the running means for the galaxies with disk features and blue cores, respectively. The FRTS (grey lines) is reproduced in the lower left corners, and the colours are the same as in Fig.~\ref{fig:vir_vs_fnx}.  \mods{The dwarf ETGs with blue colours appear to be consistent with the FRTS, and those with disk features less so.}\label{fig:dibcres}    }
\end{figure}

 In this context, it may be enlightening to analyse the early-type dwarf subtypes with disk features and blue cores as classified by \citet{2006AJ....132..497L,2006AJ....132.2432L}, which could reasonably be expected to be more recent additions to the
 Virgo cluster due to their properties and distribution relative to the cluster centre \citep{2007ApJ...660.1186L}.\footnote{\mods{ We note that \citet{2006AJ....132..497L} used dE,di as the identification of early-type dwarfs with disk features (which they identified on SDSS images, helped by unsharp masking) to emphasise the distinction from the \emph{dS0} label of \citet{1985AJ.....90.1681B}.  Furthermore, we refer the reader to \citet{2017A&A...606A.135U} for a quantitative analysis of early-type dwarfs with blue cores and for optical--near-infrared colour gradients. Images of examples of galaxies of either sub-class can be found in \citet{2006AJ....132..497L,2006AJ....132.2432L}.}}
 In Fig.~\ref{fig:dibcres}, it can be seen that those early-type dwarf galaxies with blue cores do indeed  largely follow the FRTS.\footnote{The behaviour can even be seen in the \mods{$u-g$ colour} when taking into account the two galaxies with the faintest residual surface brightnesses, which did not enter the running mean calculation.}
 However, this does not apply to the ETGs with disk features. They,\modsTwo{ on average, have} higher masses than those with
 blue cores, and several of them are more massive than the (late-type) galaxies,\mods{ which typically exhibit trends consistent with the FRTS.}
 When restricting the sample to galaxies with disk features that are less massive than $M_\star < 10^8\, {\rm M}_\odot$ (without blue cores), the residual colours do 
 not show a dependence on the residual surface brightness either. With respect to our analysis, they are more akin to the ordinary early-type dwarfs than to the blue-cored early-type dwarfs.
 This may be related to disk features being either  inherited from the progenitor or induced by an interaction in the cluster (similar to 
 multiple \modsTwo{structural} components; see \citealt{2014ApJ...786..105J}).\footnote{\mods{We note that the projected radial distributions of sub-halos and dwarf types are discussed  in detail in  \citet{2013MNRAS.432.1162L}, and for the different dwarf elliptical sub-classes in \citet{2007ApJ...660.1186L}. The distribution of ETGs and LTGs in the Fornax and Virgo clusters on the sky can be found in \citet{2019A&A...625A.143V}, their Fig.~1, and in \citet{1987AJ.....94..251B}, their Figs.~7, 9, and 10, respectively.} }

Also for the low-mass LTGs, the total range for the most extreme (residual) colours and surface brightnesses should be considered:
The span from the bluest galaxies with the highest surface brightness to those with the reddest colours and faintest surface brightnesses
 appears to be on the border of what is possible according to Fig.~\ref{fig:stellpop}. 
 However, in this figure, the evolution is only shown for ages older than 1 Gyr. Before that, the evolution
 is even faster and the tracks steeper, so that over the full range of ages, the largest differences in colour \modsL{and} surface brightness
 still seem possible. We note that in Fig.~\ref{fig:allSB}  there are some examples of blue galaxies that have bluer colours than
 any of the SSP models that we considered. These galaxies could be explained \mods{by a combination of large colour uncertainties,} younger ages, and lower metallicities.

Despite the puzzling potential inconsistency of the low-mass early types  with the FRTS, it is interesting to note that, in the mass range
in which the late types show the FRTS behaviour, the sizes of early and late types are comparable (see Fig.~\ref{fig:sizes}),
which is compatible with a scenario in which the ETGs can form via the quenching of LTGs. 
In this context, we should reiterate that the samples of both types  combined  do show slopes consistent with the FRTS at low mass.

\mods{
When comparing the two clusters, one can expect  the fraction of ETGs and the fraction of red galaxies to be sensitive to the differences in the strength of processes that may transform
galaxies from late to early type.}
However, as was noted for higher-mass galaxies \citep{1989Ap&SS.157..227F,2011MNRAS.416.1197W}, Virgo has an exceptionally 
low fraction of early-type and red galaxies. 
This has been attributed to Virgo being a dynamically young cluster that is still  in the process of forming (e.g.~X-rays; \citealt{1999A&A...343..420S};  see also \citealt{2018ApJ...865...40L}).
In the appendix (Fig.~\ref{fig:ETGfrac}), we confirm that the fraction of early-type and red galaxies of the Virgo cluster is lower than that of the Fornax 
cluster down to low galaxy masses, despite Virgo's higher mass. 
This \mods{serves as a reminder} that the Virgo cluster could quite possibly be considered as an atypical galaxy cluster. 
For example,  \mods{the analysis of \citet{2013MNRAS.432.1162L} matched the Virgo early-type dwarfs to 
 sub-halos in a dark matter simulation that fell into a large halo early (or, alternatively, lost significant amounts of dark matter after becoming a sub-halo), with the time limit chosen so that the fractions of early-type dwarfs and
 early in-fall sub-halos (those that experienced severe dark matter losses) match.
 If  \citeauthor{2013MNRAS.432.1162L}  had considered a more typical cluster, they
most likely would have found  the sub-halos  corresponding to early-type dwarf galaxies to extend
to more recent in-fall times and less extreme mass losses, simply due to the higher early-type
 fractions in \modsTwo{a} cluster more typical  than Virgo.}

While our focus was on the low-mass (late-type) galaxies for which the trends in Fig.~\ref{fig:allSB} can be interpreted as signatures of  \modsTwo{a passive stellar evolution following star formation quenching,}
it is also worth \modsTwo{mentioning} the high-mass late types, which show the opposite trend of redder colours being correlated with higher surface brightnesses for a given mass. It is reasonable to speculate that this is related, for instance, to the formation of bulges (and the evolution of bars) that lead to a concentration of evolved stars in the central parts of the galaxy.

\section{Conclusions}

We compared charts of colour versus mean effective surface brightness for galaxies binned by stellar mass for the Virgo and Fornax  galaxy clusters.
Both clusters show very similar relations, with the galaxies in the low-mass bins ($\lesssim 10^8\, {\rm M}_\odot$) having fainter surface brightnesses 
when they have red colours.
This applies for all colours probed in the wavelength coverage of the SDSS filters.
Analysing  morphological early and late types separately shows that this is found  in particular for the LTGs,
for which the slope of this relation in the lowest-mass bins is consistent with the stellar evolution model predictions for a passively
fading and reddening stellar population.
For ETGs, on the other hand, the slopes for the various mass bins are simply determined by how the colours and
surface brightnesses scale with (stellar) galaxy mass. We discussed possible explanations for the ETGs not showing
any imprints of the fading and reddening tracks even though they must have been star-forming in their past.
This discrepancy most likely has to do with an early quenching, for which the current fading and reddening becomes less significant and less well defined. 
This is supported by the Virgo early-type dwarfs with blue cores (i.e.~more recent star formation) showing clearer signs of the fading and reddening tracks.
In terms of the sizes, the ETGs and LTGs also become comparable in both galaxy clusters below a stellar mass of about $M_\star\lesssim 10^8\, {\rm M}_\odot$,
which would be consistent with a fading and reddening transformation.
\mods{While the data are insufficient to claim strong observational evidence for the expected increased strength of ram pressure in  the Virgo cluster compared to Fornax, it does not contradict this prediction.}

\begin{acknowledgements}
We thank the referee for a thoughtful reading of the draft and the detailed suggestions that helped to improve the presentation of our analysis. We acknowledge financial support from the European Union’s Horizon 2020 research and innovation program under the Marie Sk\l odowska-Curie grant agreement No. 721463 to the SUNDIAL ITN network. HS and AV are also supported by the Academy of Finland grant No. 297738. AV acknowledges the financial support from Emil Aaltonen foundation.  JJ expresses his thanks to Reynier Peletier for organizing the {\it Low Surface Brightness Galaxies} workshop in 2019 in Groningen, where initial ideas for this article originated.\end{acknowledgements}

\bibliographystyle{aa} % style aa.bst
\bibliography{vfp} % your references Yourfile.bib

\begin{appendix} %First appendix
  \section{Support material for the colour  versus surface brightness analysis}
  \label{app:sb}
\begin{figure*}
  \centering
\includegraphics[width=0.8\textwidth]{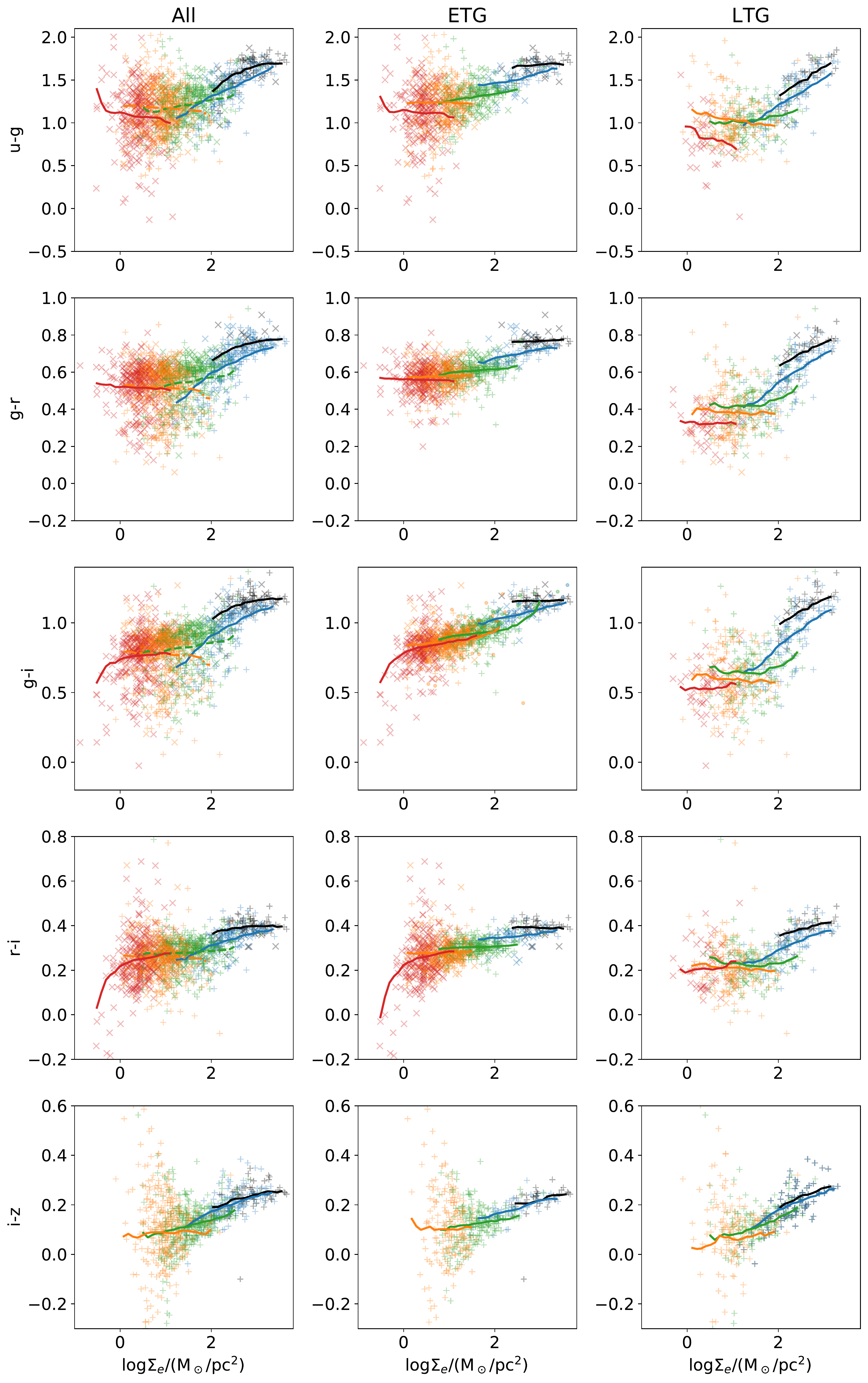}
   \caption{\mods{Colour versus mean effective mass surface density (similar to Fig.~\ref{fig:allSB}, but replacing the surface brightness with surface density);} symbols and colours are the same
   as in Fig.~\ref{fig:vir_vs_fnx}. The low-mass LTG samples that showed slopes consistent with the FRTS have flat colour relations  as a function of surface mass density, which is consistent with the interpretation of a fading and reddening stellar population. \label{fig:alldens}}
\end{figure*}

\begin{figure}
  \centering
  \includegraphics[width=0.24\textwidth]{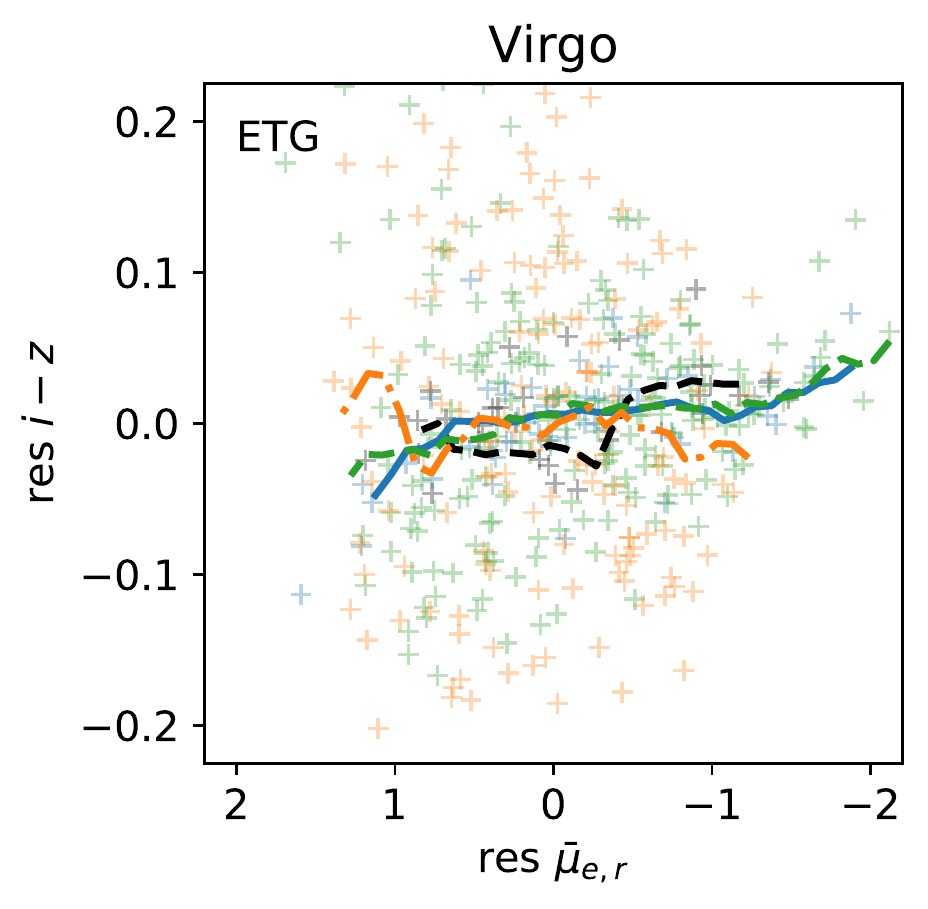}
  \includegraphics[width=0.24\textwidth]{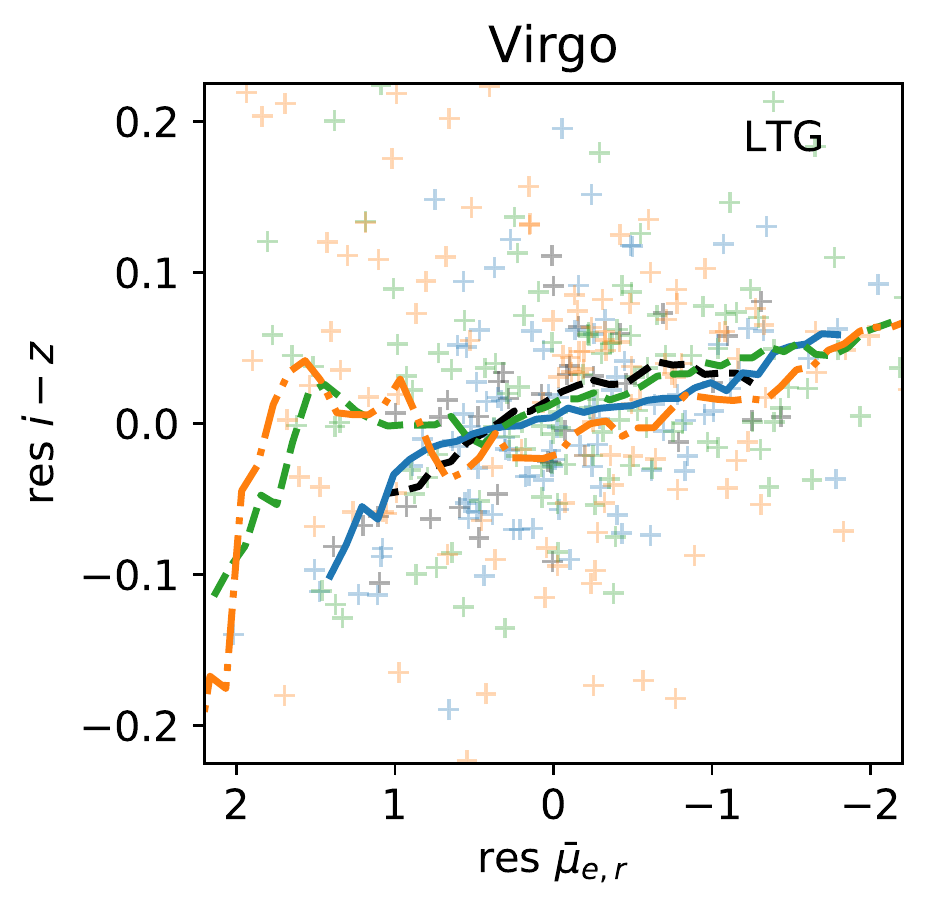} 
 \caption{\mods{Extra panels for Fig.~\ref{fig:ETGres} with  $i-z$ colour for the Virgo cluster for ETGs \emph{(left)} and LTGs \emph{(right)}. The $i-z$ colour is noisier, and the slopes are consistent with being flat.}\label{fig:additional}}
 \end{figure}

\begin{figure}
  \centering
     \includegraphics[width=0.34\textwidth]{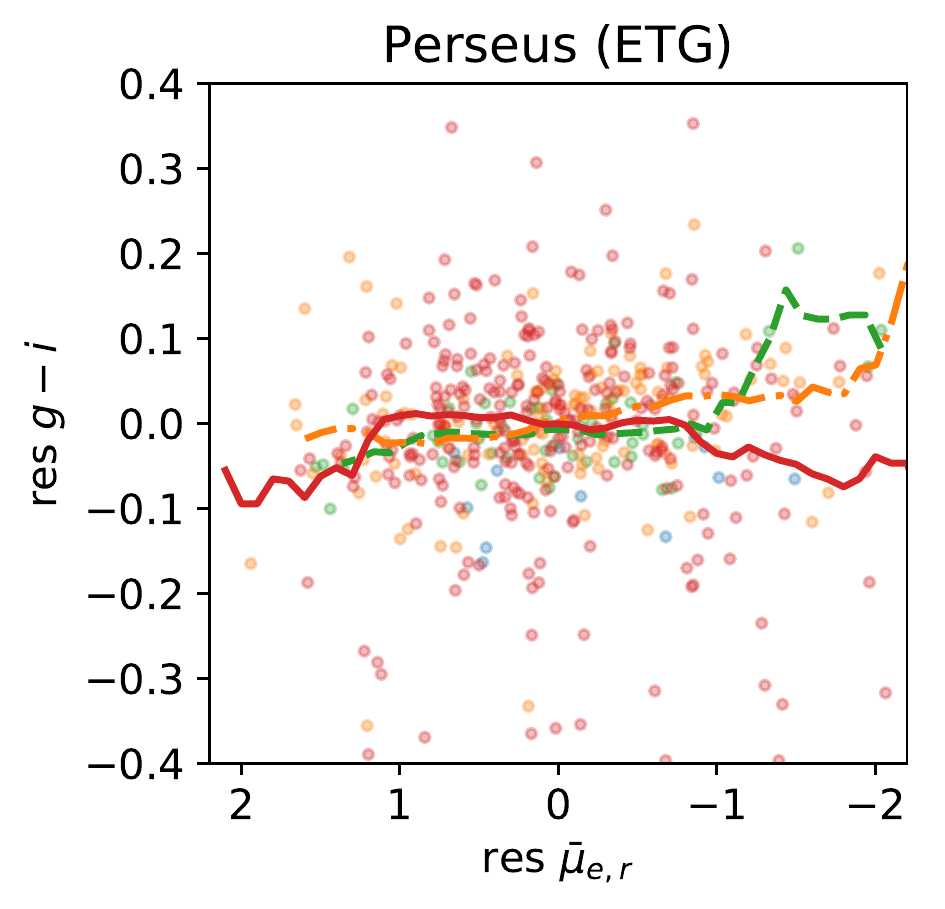}
\caption{\mods{Residual colour versus residual surface brightness relation for ETGs in the Perseus cluster, as in Fig.~\ref{fig:ETGres}.
The relations are consistent with the observations in the Virgo and Fornax clusters.
\label{fig:perseus}}}
\end{figure}

\begin{figure}
  \centering
     \includegraphics[width=0.495\textwidth]{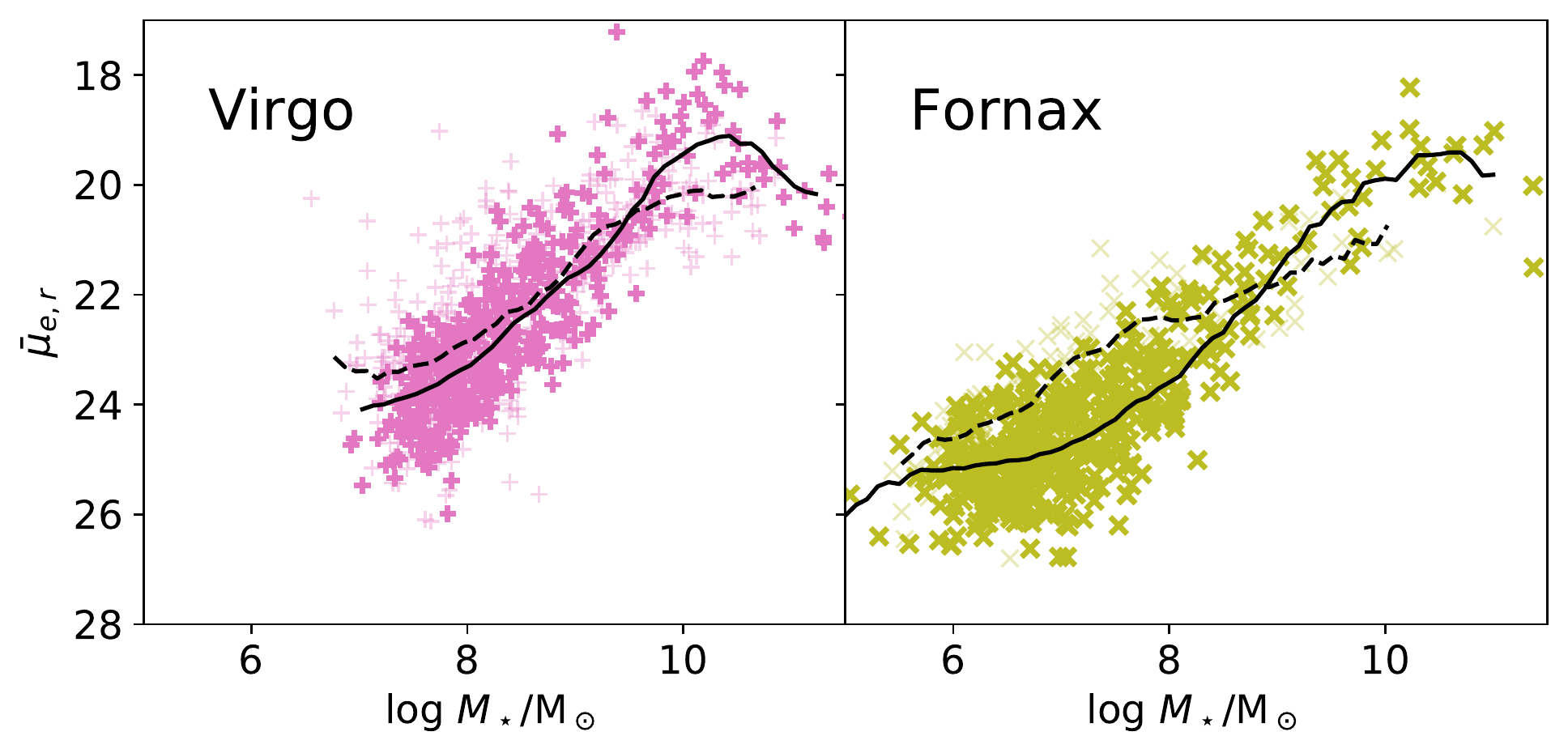}
\caption{\mods{Mean effective surface brightness versus stellar mass; symbols are the same as in Fig.~\ref{fig:cmr}.
These average relations (solid curves -- ETGs, dashed curves -- LTGs) were used to calculate the residual colour versus mean effective surface brightness relations in Figs.~\ref{fig:ETGres} and \ref{fig:LTGres}. They are the running means with a window width of 1 dex.
\label{fig:SB}}}
\end{figure}

In this appendix, the following support material for the analysis is provided. Figure~\ref{fig:alldens} is similar to Fig.~\ref{fig:vir_vs_fnx} but plots effective mean stellar surface density instead of surface brightness. Figure~\ref{fig:additional} shows additional panels for the residual colour versus surface brightness plot (Fig.~\ref{fig:ETGres}) for Virgo in $i-z$. Figure~\ref{fig:perseus} is the corresponding figure for the ETGs in the Perseus cluster.
Finally, Fig.~\ref{fig:SB} shows the running means of the surface brightness \modsTwo{versus stellar mass} used for the residual colour versus surface brightness diagrams.

The stellar surface densities are obtained from the surface brightnesses and colours using \citet{2010ApJ...722....1T}. The flat curves for low-mass LTGs in Fig.~\ref{fig:alldens} support the interpretation \modsL{that} they are the result of passively fading and reddening stellar populations since in this picture the stellar densities do not change.

\mods{ Figure~\ref{fig:additional} shows the residual relations for the Virgo galaxies with the $i-z$ colour that was omitted from Figs.~\ref{fig:ETGres} and \ref{fig:LTGres}  due to the lack of corresponding data for the Fornax cluster.
In Fig.~\ref{fig:alldens}, data for ETGs of the Perseus cluster from \citet{2017MNRAS.470.1512W} were added to the early-type panel with $g-i$ colours. The $V-I$ colours were converted to $g-i$ using the transformations provided in \citet{2006A&A...460..339J}.}
The Perseus cluster is about four times more massive than Virgo and more than 11 times more massive than Fornax (using the fiducial mass estimates in \citealt{2011MNRAS.416.1197W}).
Also for this cluster, the slopes of ETGs in different mass bins in the colour surface brightness plots resemble one another.
The corresponding plot with the residual quantities after subtraction of the colour and surface brightness scaling relations is shown in \mods{Fig.~\ref{fig:perseus}.} As for the other clusters, the behaviour of the ETGs is a result of the scaling relations.\mods{
In Fig.~\ref{fig:SB}, the running mean relations for the mean effective surface brightness as a function of stellar mass, which were used to calculate the residual colour versus residual surface brightness relations, are displayed for reference. }

\section{FRTS and more complex star formation histories }
\label{app:csp}
\mods{
The slopes (FRTS) of the fading and reddening tracks used throughout our analysis are listed in Table~\ref{tab:slopes}. They were obtained through linear fits of the fading and reddening tracks for ages from 2 to 12 Gyrs  in Fig.~\ref{fig:stellpop}. The slopes do not depend strongly on the metallicity of the stellar population.}

\mods{
\begin{table}
\caption{\mods{Slopes of the relations fitted to the fading and reddening tracks ($\Delta$colour$/\Delta$m).\label{tab:slopes}}}
\begin{center}
\begin{tabular}{cccccc}
[M/H] &$u-g$ & $g-r$ & $g-i$ & $r-i$ & $i-z$ \\
\hline
-0.96 & 0.1230 & 0.0472 & 0.0724 & 0.0252 & 0.0082 \\
-0.66 & 0.1318 & 0.0520 & 0.0787 & 0.0267 & 0.0111 \\
-0.35 & 0.1403 & 0.0588 & 0.0879 & 0.0291 & 0.0171 \\
-0.25 & 0.1402 & 0.0588 & 0.0875 & 0.0287 & 0.0179 \\
0.06 & 0.1378 & 0.0580 & 0.0851 & 0.0272 & 0.0199 \\
\end{tabular}
\end{center}
\end{table}
}

\mods{
As described in the text, the applicability of the SSP models to the LTGs with typically extended star formation histories needs to be tested. In order to address this question, we examined the FRTS for two alternative star formation histories.
One type of star formation history that is typically used to describe LTGs is exponentially declining models (for which the star formation rate initially ramps \modsTwo{up} according to a power-law).
We are interested in the evolution of the brightness and colour 
of these models after the star formation is truncated. As such, the star formation rate was set to 0 at a time $t_{\rm trunc}$: 
\begin{equation}
 SFR(t)=\left\{
        \begin{array}{ll}
            \left({{t_{\rm start} - t}\over{t_{\rm start}}}\right)^n \times \exp{\left[-\frac{t_{\rm start} - t}{\tau}\right]} & t > t_{\rm trunc} \\
            0 &   t < t_{\rm trunc}
        \end{array}
    \right.
.\end{equation}
\modsTwo{Here, $t$ corresponds to the lookback time, $\tau$ corresponds to the time scale of the exponential decline, and the start of the star formation, $t_{\rm start}$, is set to $13$ Gyr.}
This type of model was suggested by, for example, \citet[their choice to model an Sc galaxy with truncated star formation]{1999ApJ...518..576P}.
We used the SSP models to approximate these more complex star formation histories using a large number of discrete bursts (we used 100, which we spaced logarithmically so as to better sample the more rapid evolution at younger ages). }

\mods{For an individual galaxy, the abovementioned exponentially declining star formation rate may or may not be a good approximation and for a population of galaxies the parameters will depend on, for example, the mass of the galaxy. We do not try to model all of these complexities. Instead, we used a second type of star formation history, with a constant star formation rate, for the following demonstration; it corresponds to another extreme (the instant star formation of the SSPs being the first extreme and the more realistic exponentially declining star formation law in between the two):}
\begin{equation}
 SFR(t)=\left\{
        \begin{array}{ll}
            const & t > t_{\rm trunc} \\
            0 &   t < t_{\rm trunc}
        \end{array}
    \right.
.\end{equation}

\begin{figure}
  \centering
\includegraphics[width=0.5\textwidth]{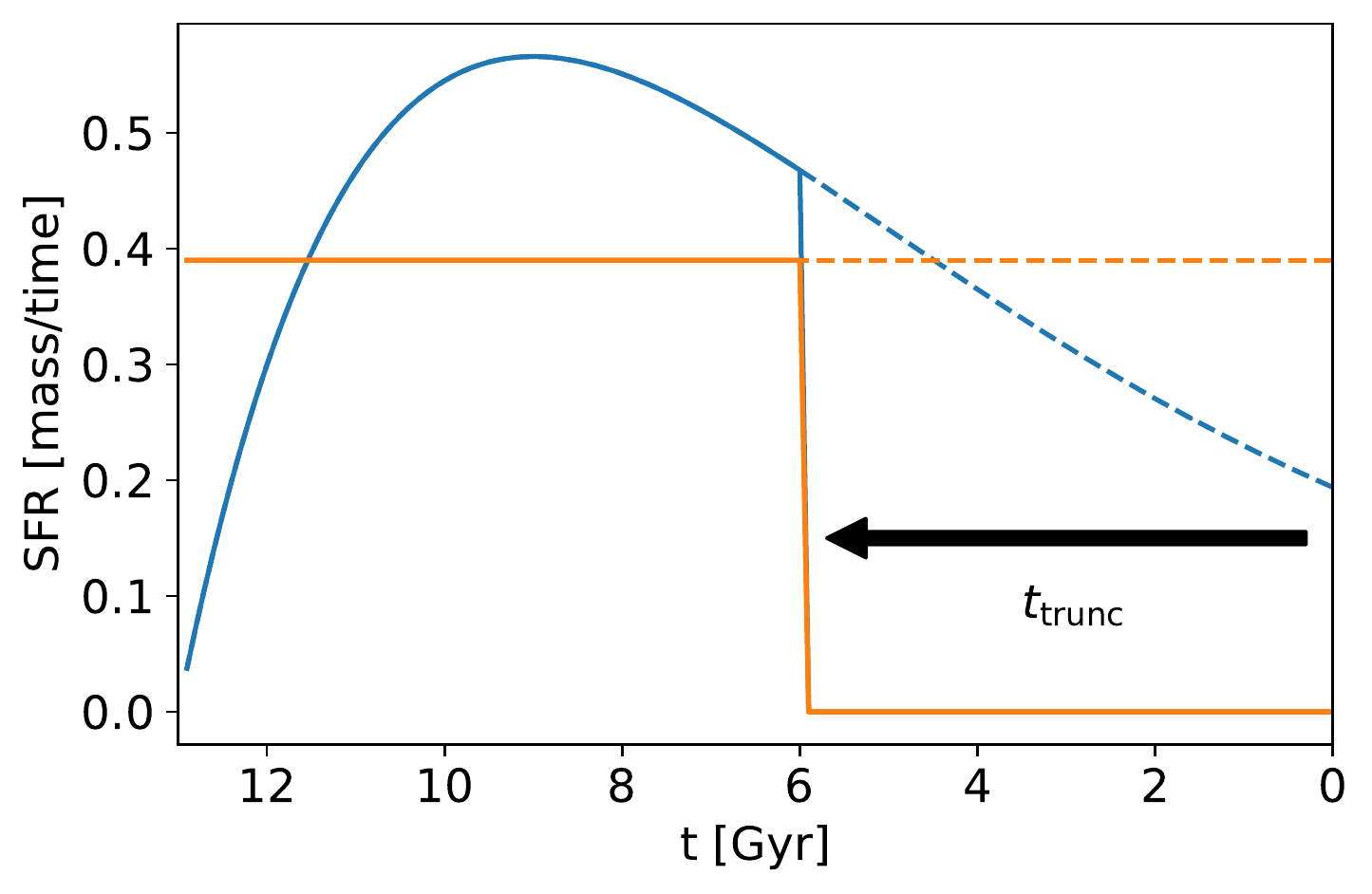}
   \caption{\mods{Illustration of the considered (complex) star formation histories: an exponentially declining star formation rate (blue) and a constant star formation rate, each truncated by the interaction with the cluster at $t_{\rm trunc}$. \modsTwo{Here the times correspond to lookback times, and $\tau=4$ Gyr for the exponentially declining model.} An arbitrary normalisation is applied to the y-axis for better readability. 
   \label{fig:sfr}}}
\end{figure}

\mods{Both types of star formation histories are illustrated in Fig.~\ref{fig:sfr}.
 In order to account for the binning by present-day mass, we normalised the truncated models so that they form the same amount of stars by the truncation time $t_{\rm trunc}$. We note that the curves representing the models in colour versus brightness diagrams no longer represent fading tracks of an individual galaxy but instead describe how the colours and magnitudes of different galaxies of the same mass, with the same form of star formation history, but truncated at different times, are related. }
\mods{
By showing that both models --  an example for an exponentially declining star formation (with $t_{\rm start}=13$ Gyrs, $\tau=4$ Gyrs, $n=1$)  and  that with a constant star formation rate -- give slopes in the colour-(surface) brightness diagram consistent with those of the SSPs, we demonstrate that the details of the star formation prior to the quenching are not very significant for our analysis.\footnote{We note that this ignores any changes of the metallicities throughout a galaxy's history due to enrichment by evolved stars and inflows of pristine gas, which would need to be considered in full models.}}

\mods{There is one exception to the above statement: models with a  short time scale for the exponential decline $\tau$ (i.e. when the majority of stars form well  before the considered $t_{\rm trunc}$). While the slope does not dramatically change in this case either, the total difference of colour and brightness between early and late truncation times shrinks considerably. This is consistent with the discussion regarding an early  formation for the stars in ETGs as a way to explain their lack of FRTS behaviour. It should  be said that such a scenario is, however, somewhat at odds with the idea of downsizing (i.e.~longer time scales for the star formation in low-mass galaxies; \citealt{1996AJ....112..839C}) and the observations of Local Group early-type dwarfs, which are observed to have very similar star formation histories as their late-type counterparts, when ignoring the star formation at most recent times (which is reflected in their classifications as late or early types; see \citealt{2014ApJ...789..147W}).}

\begin{figure*}
  \centering
\includegraphics[width=\textwidth]{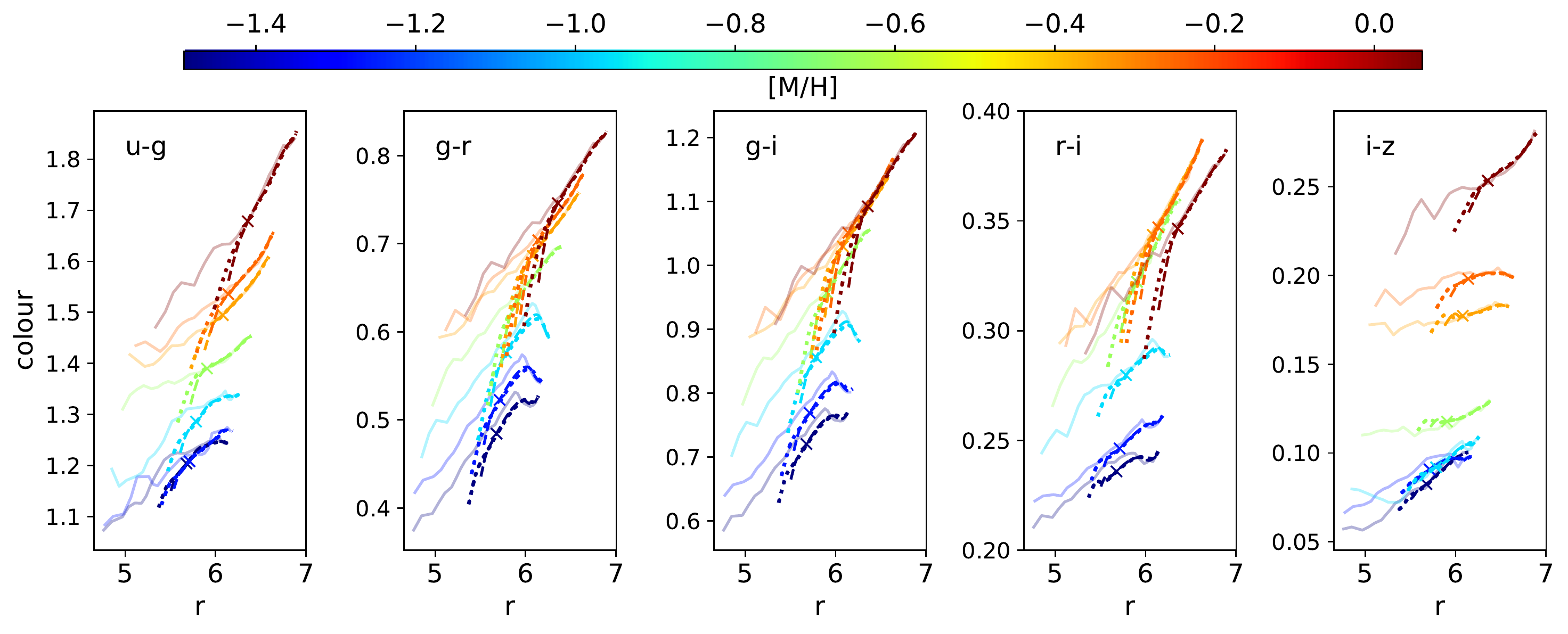}
   \caption{\modsThree{Comparison of colours and brightnesses of galaxies with complex stellar populations \modsL{(CSP;} see Fig.~\ref{fig:sfr}) with the same present-day stellar mass but different truncation times. This figure, for the CSP, corresponds to Fig.~\ref{fig:stellpop} for the SSP models.} The solid pale lines reproduce these SSP relations for ages between 2 and 12 Gyrs, the range over which the slopes have been fitted. The dotted and dashed lines show the curves for models with more complex star formation histories (constant and exponentially declining, respectively; see text) with truncation times between 0.5 and 12 Gyrs ago. The resulting slopes are largely consistent with those of the SSP, with a steeper slope at truncation ages younger than 2 Gyrs (marked with  crosses).\label{fig:stellpopCSP}}
\end{figure*}

\section{Further Virgo versus Fornax comparisons}
\label{app:FV}
\begin{figure}
  \centering
  \includegraphics[width=0.45\textwidth]{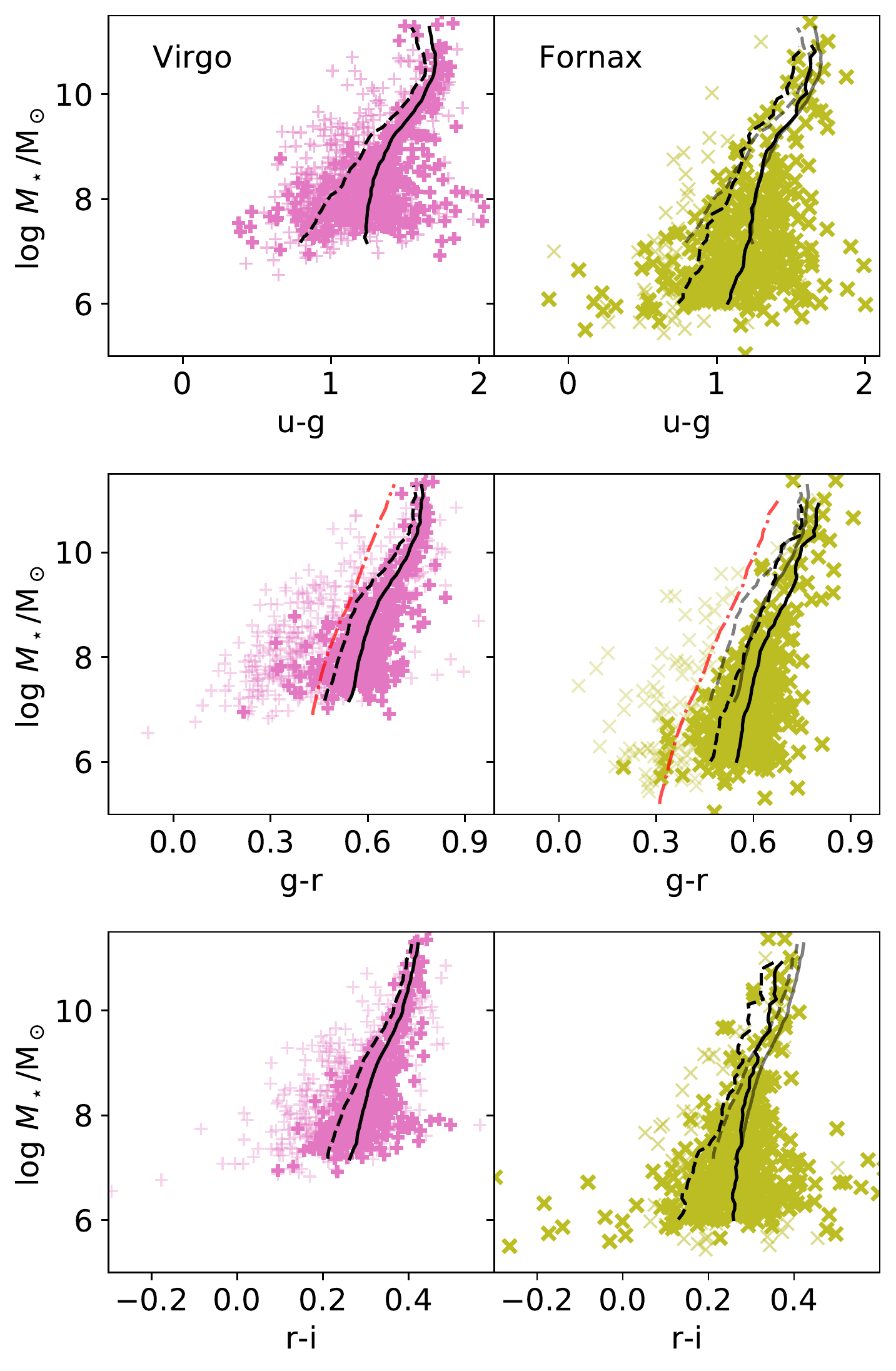}
    \caption{As in Fig.~\ref{fig:cmr}, stellar mass versus colour relations for galaxies in the Virgo \emph{(left)} and Fornax
   \emph{(right)} clusters. The red curve in the panels with $g-r$ illustrates the dividing line for the red fraction, according to \citet[\modsTwo{to illustrate their linear relation as a function of total magnitude in this figure,} the stellar masses were converted to $r$-band brightnesses using a running mean average relation]{2011MNRAS.416.1197W}. The relations are largely consistent for both clusters.
  \label{fig:cmrall}          }
  \end{figure}
  In this appendix, we show further comparisons between the Virgo and Fornax clusters.
In Fig.~\ref{fig:cmrall}, we extend the comparison of the colour-mass relation shown \modsL{at the beginning of the paper} to the other colours.
As in Fig.~\ref{fig:cmr}, it can be seen that the small offset in the mean colours of ETGs
at a given mass between the two clusters depends on the stellar mass.

\begin{figure*}
  \centering
\includegraphics[width=0.9\textwidth]{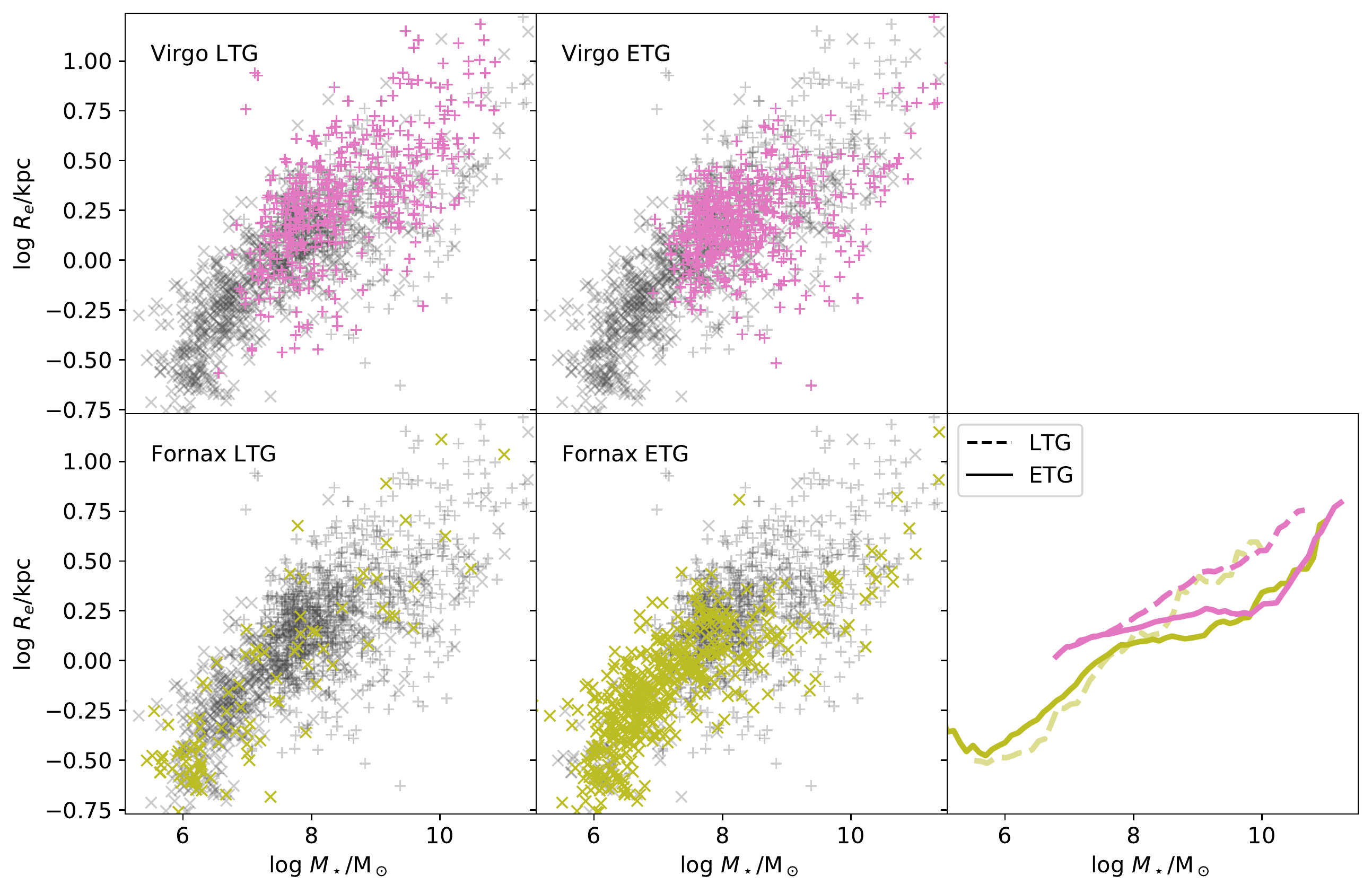}
   \caption{Half-light size (semi-major axis lengths) versus stellar mass for
   LTGs and ETGs in the Virgo and Fornax
   clusters. Each panel with data points shows one highlighted sub-sample
  \mods{ (see the label in the panel), and all other galaxies are shown with grey symbols (pluses -- Virgo, crosses -- Fornax).} The rightmost panel shows the running mean relations as curves. \mods{The size--mass relations are consistent for both clusters, and at low galaxy masses the
   sizes of LTGs and ETGs become comparable in both clusters.}
\label{fig:sizes}           }
\end{figure*}

 Figure~\ref{fig:sizes}  compares the size versus mass scaling relation of ETGs and LTGs in the Virgo and Fornax clusters.
 The mean relations for each morphological type as quantified by the running means\mods{ are consistent  for both clusters.}  At masses $M_\star \lesssim 10^8\, {\rm M}_\odot$, the sizes of ETGs and LTGs in each cluster match well, which is consistent with a simple fading and reddening evolution.

  \begin{figure}
     \centering
   \includegraphics[width=0.48\textwidth]{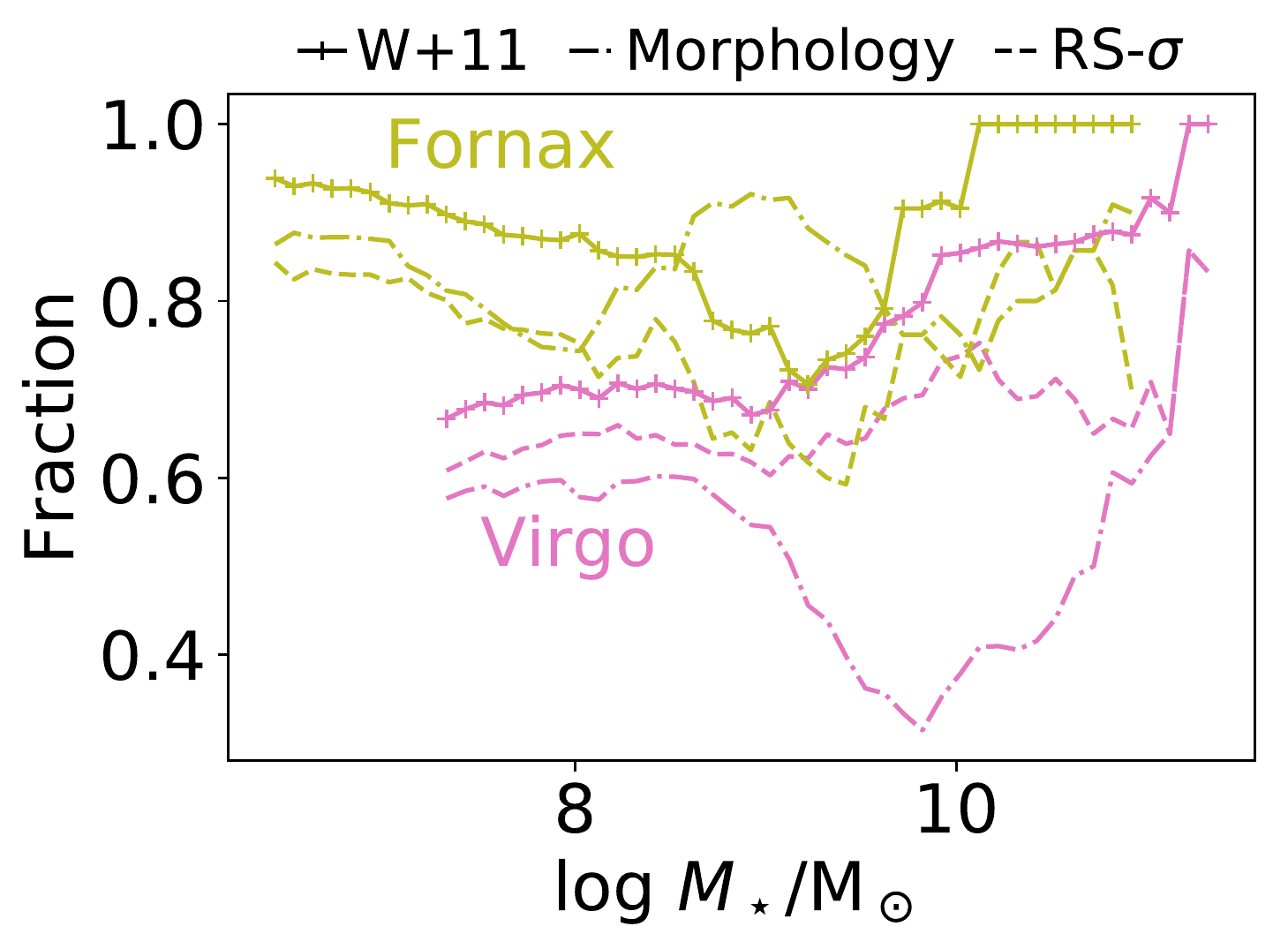}
    \caption{Fraction of red \modsL{and} early-type galaxies in the Virgo \emph{(magenta)} and Fornax \emph{(green)} clusters.
    The lines show the red or early-type fraction according to different definitions.
        The W+11 line corresponds to a linear demarkation line as a function of  total magnitude bluewards of the red sequence from
    \citet{2011MNRAS.416.1197W}. The second, dash-dotted line is for the early-type fraction according to the morphological 
    classification. The RS-$\sigma$ line corresponds to a split between the red and blue galaxies\mods{ along the red sequence \modsL{(RS)} traced by running means minus the $1\sigma$ scatter in the running bins of 1 dex width (see the dashed lines in Fig.~\ref{fig:cmr}).  The Fornax cluster has, by all measures, a higher fraction of early-type and red galaxies throughout almost the whole range of stellar mass.}
   \label{fig:ETGfrac}              }
   \end{figure}

The Virgo cluster is considered \mods{to be a} dynamically young, less evolved cluster that is possibly
undergoing an active assembly phase. This is reflected in the various subgroups, offset from the centres defined by
the galaxy distribution versus the dominant galaxy and peak of X-ray emission, by the double peaked distribution
of X-ray emission, and, last but not least,
by an unusually high fraction of late-type and star-forming galaxies.
In Fig.~\ref{fig:ETGfrac},   we compare the fraction of ETGs and the fraction of red galaxies in the Virgo and Fornax
 clusters as a function of galaxy mass. The morphological classification indeed shows the higher late-type fraction of the
 Virgo cluster at all galaxy masses. Since the Fornax morphological classifications are a simple division of early versus late type, and since the classifications were done \mods{by different individuals,} we also show the fraction of red galaxies.
 Here, we used the fixed boundary between red and blue galaxies with a linear slope as a function of brightness defined by \citet[$g-r > 0.4 - 0.03 \times (M_r+13)$; Fig.~\ref{fig:cmrall}]{2011MNRAS.416.1197W}\mods{ and, alternatively, one that is based on the running mean relation minus the 1$\sigma$
scatter (see the dashed line in Fig.~\ref{fig:cmr}) in the running bins (1 dex width). All of the curves consistently show a lower early-type \modsL{or} red  fraction for the Virgo cluster at all galaxy
masses despite its higher cluster mass.}

\end{appendix}
\end{document}